\documentclass[12pt,reqno]{amsart}
\usepackage{a4wide}
\usepackage{amssymb}
\usepackage{enumerate}{}
\newcommand{\e}{\varepsilon}

\newcommand{\R}{\mathbb{R}}
\newcommand{\TT}{\mathbb{T}}

\let\a=\alpha

\let\d=\delta
\let\e=\varepsilon

\let\G=\Gamma

\newcommand{\1}{\,\rlap{\small 1}\kern.13em 1}

\newcommand{\sqr}[2]{{\vcenter{\hrule height.#2pt%
                      \hbox{\vrule width.#2pt height#1pt\kern#1pt%
                            \vrule width.#2pt}%
                      \hrule height.#2pt}}}

\let\e=\ve

\newtheorem{thm}{Theorem}[section]

\newtheorem{coro}{Corollary}[section]

\begin{document}

\title[]{Phase segregation and interface dynamics \\ in kinetic systems.}

\author{Guido Manzi$^\dag$}

\author{Rossana Marra$^*$}

\dedicatory{Dipartimento di Fisica, Universit\`a  di Roma Tor Vergata e INFN\\ Via della Ricerca
Scientifica 1
 00133 Roma, Italy\\$^\dag$ manzi@roma2.infn.it\\$^*$ marra@roma2.infn.it}


\begin{abstract}
 We consider a kinetic model
 of two species of particles interacting with a reservoir at fixed temperature, described by two coupled
Vlasov-Fokker-Plank equations. We prove that in the diffusive limit the evolution is described by a macroscopic
equation in the form of the gradient flux of the macroscopic free energy functional. Moreover, we  study the
sharp interface limit and find by formal Hilbert expansions that the interface motion is given in terms of a quasi
stationary problem for  the chemical potentials. The velocity of the interface is the sum of
two contributions:  the velocity of the Mullins-Sekerka motion for the difference of the chemical potentials and
the velocity of a Hele-Shaw motion for  a linear combination  of the two potentials. These
equations are identical to the ones in
\cite{OE} modelling the motion of a sharp interface for a polymer blend.
\end{abstract}

\maketitle


\bigskip

\section{Introduction}

When a fluid mixture is suddenly quenched from a homogeneous equilibrium state into a thermodynamically unstable
state it evolves to an equilibrium state consisting of two coexisting phases, each one reacher in one species. This
 phenomenon is called phase segregation. There  are various stages during  this process, starting with the
formation of interfaces very diffuse, that sharpen  with time  and then move slowly driven by surface tension
effects. In \cite{BELM1}  the so-called late stages of the phase segregation process have been investigated for
a kinetic model of
a  system of two species of particles interacting by a repulsive
long range potential and collisions.  The repulsive  interaction between different species is modeled by a Vlasov
term and the collisions by a Boltzmann Kernel. The evolution of the system is then ruled by two coupled
Vlasov-Boltzmann equations
 for the one-particle distributions  and this dynamics conserves masses momentum and energy. In
the late stages of the coarsening process the hydrodynamics effects in this case become relevant and when the
fluid is well segregated with sharp interfaces between different phases the interface moves in its normal
direction following the incompressible velocity field solution of the Navier-Stokes equation, while the pressure
satisfy the Laplace's law relating it to the surface tension and curvature \cite{BELM1} .
In the present paper, we study   a similar   kinetic model replacing the Boltzmann kernel by
a Fokker-Plank operator, namely    the two species interact with the
same reservoir at fixed temperature instead of colliding each other, so that only the masses are conserved. We are
in a situation in which the temperature and the momentum relax to equilibrium  faster than the densities, as in the
polymer blends where the viscosity is very large. This  model can be seen as the kinetic description of a system  of
particles interacting via a weak long range potential (and with a reservoir)  in the real space (as opposite to the
lattice)  or as the mean field limit of a stochastic system of interacting particles on the continuum
(Ornstein-Ulhenbeck interacting processes).   System of particles on the continuum are more difficult than the
corresponding ones on the lattice because of the control of the local number of particles: the conservation law
cannot prevent locally very high densities. Systems of this kind on the lattice  have been introduced in a series of
papers  \cite {GLP} (and references therein) to study phase separation in alloys  and their behavior has been widely
investigated.The macroscopic evolution of the conserved order parameter is ruled by a nonlinear nonlocal integral
differential equation having non homogeneous stationary solutions at low temperature, corresponding to the presence
of
 two different phases separated by interfaces. When the phase domains are very large compared to
the size of the interfacial region (so-called  sharp interface limit) the interface motion is
described in terms of a Stefan-like problem or the Mullins-Sekerka motion depending on the time
scale \cite {GL}. In this paper we derive rigorously macroscopic equations for the density profiles and study
by formal expansions the sharp interface limit.

 The equations for the one-particle distributions $f_i(x,v,\tau)$
are
\begin{equation}\partial_{\tau} f_{i} +v\cdot \nabla_xf_{i} +F_{i}\cdot \nabla_v
f_{i} =L_\beta f_{i} \quad i ,j=1,2, \quad i\neq j, \label{0.2}
\end{equation}
where $\beta$ is the inverse temperature of the heat reservoir modeled by the Fokker-Plank
operator on the velocity space $\mathbb{R}^3$ 
$$
L_\beta
f_{i}:=\nabla_v\cdot\biggl(M_\beta\nabla_v\biggl(\frac{f_i}{M_\beta} \biggr)\bigg),\quad
 M_\beta(v) = (\frac{2\pi}{\beta } )^{-3/2}\exp(-\beta|v|^2/2)
 $$
and $F_{i}$ are the self-consistent forces, whose potential has inverse range $\gamma$, representing the repulsion
between particles of different species: 
\begin{equation}
F_{i}(x,\tau)= -\nabla_x\int {\rm d}\/x' \gamma^3 U(\gamma|x-x'|)\int
{\rm d}vf_{j} (x',v,\tau),\quad i,j=1,2, \quad i\neq j.\label{0.3}
\end{equation}
Our system is contained
in  a $3$-dimensional torus  (to avoid boundary effects) and $U(r)$ is a non negative, smooth
monotone  function on
$\mathbb{R}_+$ with compact support. This evolution conserves the  total masses   of the two
species.
 Beyond the spatially constant equilibria, there may be other spatially
non homogeneous stationary solutions.  To characterize the stationary solutions is useful to consider the
coarse-grained functional $ \mathcal G$
\begin{eqnarray*}
\mathcal G (f_1,f_2) := &\displaystyle{ \int {\rm d }x {\rm d }v[(f_1\ln f_1)(x,v) + (f_2\ln
f_2)(x,v)]+\frac{\beta}{2}\int {\rm d }x {\rm d }v(f_1+f_2) v^2 }&\\ &\displaystyle{ +\beta\int {\rm d}x{\rm
d}y\gamma^3 U(\gamma(x-y))\int {\rm d }v f_1(x,v)\int {\rm d
}v' f_2(y,v')}\\ \label{freeg}
\end{eqnarray*}
It is easy to see that $\mathcal G$ is  a  functional decreasing in time under the Vlasov-Fokker-Plank dynamics. In
fact,  only the Fokker-Plank  term  gives a contribution different from zero to  the time derivative of $
\mathcal G$ which satisfies
\begin{equation}
\frac{d}{dt}\mathcal G= \sum_{i=1,2}\int {\rm d}x{\rm d}v\nabla_v\cdot ( M_\beta
\nabla_v\frac{f_{ i}}{M_\beta})\ln \frac{f_i}{M_\beta}=-\sum_{i=1,2}\int {\rm d}x{\rm d}v\frac
{{M_\beta}^2}{f_i} [\nabla_v\frac{f_i}{M_\beta}]^2\le 0
\end{equation}
and it is zero only and only if ${f_i}$ are Gaussian functions of the velocity. Hence,
the equilibrium states  are local Maxwellian
with mean value $u=0$, variance $T=\beta^{-1}$, and
densities $\rho_i=\int dv f_i(x,v,\tau)$ satisfying
\begin{equation}
T\log \rho_i(x)+\int {\rm d}x' \gamma^3U(\gamma |x-x'|) \rho_j(x')=C_i, \quad i=1,2, i\neq j.
\label{0.1}
\end{equation}

Moreover,  the functional $\mathcal G$ evaluated on functions of the form $f_i(x,v,\tau)=M_\beta(v)\rho_i(x,\tau)$
with fixed total masses $\int {\rm d}vf_i$, coincides apart a constant  with
the
 macroscopic
free energy functional
\begin{equation}
{\mathcal F}(\rho_1,\rho_2) = \int{\rm d }x\left[ (\rho_1\ln\rho_1)(x) +
(\rho_2\ln\rho_2)(x)\right] + \beta\int {\rm d}x{\rm d}y\gamma^3
U(\gamma(x-y))\rho_1(x)\rho_2(y)\ . \label{free)}
\end{equation}
It is proved in \cite{CCELM}, under the assumption of a monotone potential, that at low temperature there are non
homogeneous solutions to (\ref{0.1}), stable in the sense that they minimize ${\mathcal F}$.  On the infinite
line the non homogeneous minimizers of the excess free energy  under fixed asymptotic values are called fronts and
have monotonicity properties.  The asymptotic values at $\pm\infty$ are the values  of the densities of two
coexisting different phases at equilibrium, one reach in species $1$ and the other reach in species $2$.

The macroscopic equations, which play the role of the Cahn-Hilliard equations for this model,
are obtained in the diffusive limit: they describe the behavior of the system on length scales of order $\e^{-1}$
and time scales of order $\e^{-2}$ in the limit of vanishing $\e$, where $\e$ is the ratio between the kinetic and
the macroscopic scale. Moreover, we choose $\gamma=\e$ so that the range of the potential is finite on the
macroscopic scale. We prove in sect.s 2,3,4 that in this limit solutions of (\ref{0.2}) converge to solutions of
the  following coupled non local parabolic equations for the densities $\rho_i(x, t)$
\begin{equation}\beta^2\partial_{ t} \rho_{i}=\Delta \rho_i+
\beta\nabla\cdot(\rho_i\nabla U\star
\rho_j),\quad i ,j=1,2,\ i\neq j \label{0.4}
\end{equation}
where $(U\star g)(x, t)=\int {\rm d}y U(x-y) g(y, t)$. These equations can
be rewritten in the form of a gradient flux for  the  free energy functional ${\mathcal F}$
\begin{equation}
\partial_{ t} \bar {\rho}=\nabla\cdot \Bigg({\mathcal M}\nabla \frac{\delta{\mathcal
F}}{\delta\bar {\rho}
}\Bigg),\quad {\mathcal M}=
\beta^{-1}
\begin{Bmatrix}\rho_1&0\cr 0&\rho_2\cr\end{Bmatrix}
\label{0.5}
\end{equation}
where $\bar {\rho}=(\rho_1,\rho_2)$, $\frac{\delta{\mathcal F}}{\delta \rho_i}$
denotes the
functional derivative of
${\mathcal F}$ with respect to $\rho_i$ and
${\mathcal M}$ is the $2\times 2$ mobility matrix.
This form of the equation is very important to study the stability
properties of the stationary solutions. Since we know that the stationary solutions are
minimizers of the functional ${\mathcal F}$, we expect to be able to prove that  the system relaxes to
that stationary state asymptotically in time, for example using the approach developed in
\cite{DOPT} for a nonconservative equation.

To describe the late stages of the segregation process,  we scale position and time as $\e^{-1}$ and $\e^{-3}$
respectively, while keeping fixed (equal to $1$) $\gamma$ in the Vlasov-Fokker-Plank equations (\ref{0.2}).
 The width of the interface on the macroscopic scale is then of order $\e$, so that in the
limit $\e\to 0$ the interface becomes sharp. On the same scales of space and time the motion of the interfaces for
models of alloys is given by the Mullins-Sekerka model \cite{MS}, a quasi stationary boundary problem
in which the mean curvature of the interface plays a fundamental role. We find similar results in our case, but
with relevant differences. We choose an initial condition with an interface $\G_0$ and profiles for the densities
given by
 front solutions with
asymptotic values $\rho_1^\pm=\rho_2^\mp=\bar\rho^\pm$
 corresponding to the equilibrium values in the phase transition region at temperature $T$. In the limit $\e\to
0$ the difference of the first correction to the chemical potentials $\mu_i$,
 $\psi=\mu_1^{(1)}-\mu_2^{(1)}$ satisfies
\begin{equation}
\left\{
    \begin{array}{ll}
     \Delta_r\psi(r, t) = 0\qquad\quad \mbox{for $r\in \Omega\setminus\Gamma_t$} &\\   \displaystyle{\psi(r,t)
= \frac{S K(r,t)}{\bar\rho^+-\bar\rho^-}  \qquad\mbox{$r\in \Gamma_t$}}&\\ V
=\frac{T}{2[\bar\rho^+-\bar\rho^-]}\Big[\frac{1}{{\bar\rho}}
(\bar\rho^2-|\bar\varphi|^2){\left[\nu\cdot\nabla_r\psi\right]_-^+} +
|\bar\varphi|\nu\cdot\nabla_r\zeta\Big]&\\
\end{array}
\right .
\end{equation}
where $[ \ ]^+_- $ denotes the jump across the interface $\Gamma_t$, $\bar \rho$ and  $\bar\varphi$ are the values of
total density and concentration respectively at equilibrium,
$K$ is the  curvature in $r$ of $\Gamma_t$ (sum of principal curvatures) and $S$ the surface tension for this model
(see Appendix B). This is similar to the Mullins-Sekerka equation  but for the fact that there is an extra term
determining the velocity   proportional to
 $ \nu\cdot\nabla_r\zeta(r,t), r\in \Gamma_t$  where  $\zeta(r,t)
=(\bar\rho_1\mu_1^{(1)}+\bar\rho_2\mu_2^{(1)})(r,t)$ is solution of
\begin{equation}
\left\{
    \begin{array}{ll}
     \Delta_r\zeta(r, t)=0\qquad\quad\mbox{for $r\in \Omega\setminus\Gamma_t$}&\\  \displaystyle{[\zeta]^+_-
=2|\bar\varphi| S K(r,t)  \qquad\mbox{$r\in \Gamma_t$}}&\\
  0 =\left[\nu\cdot\nabla_r\zeta\right]_-^+&
    \end{array}
\right.
\label{HS}
\end{equation}
The extra term is the normal interfacial velocity in the Hele-Shaw interface motion (\ref{HS}). The equations for
$\psi$ and $\zeta$ are identical to the ones in \cite{OE}, describing the interface motion of an incompressible
fluid mixture   driven by thermodynamic forces, modeling a polymer blend. A discussion on
that point is in sec. 7.
 In sec. 6 we study the sharp interface limit by means of formal expansions of the Hilbert type.

As last remark, we observe that the usual approach in literature to study the sharp interface  limit  is to 
start from the
 macroscopic equations (e.g. Cahn-Hilliard) and send to zero the ratio between
 the width of the interfacial region and the linear size of the phase domains.  That has been
studied  in \cite{MM} for  the macroscopic equations
(\ref{0.4}) by means of formal matching expansions.
\bigskip

\section{Macroscopic limit:expansion}
\setcounter{equation}{0}

In this section we begin the study of the hydrodynamical limit for the Vlasov-Fokker-Plank equations (\ref{0.2}).
We consider   the diffusive scaling in  which the space is scaled as
$\varepsilon$ and time as $\varepsilon^2$ and $\gamma=\e$ so that the width of the interface is of order $1$
on the macroscopic scale. Define
\[
{f_i^\e}(x,v,t):=f_i(\varepsilon^{-1}x,v,\varepsilon^{-2}t), \qquad  i=1,2,  \ x\in \TT^d, v\in
\mathbb{R}^d.
\]
The equation for ${f_i^\e}$ is
\begin{equation}\label{eq-res}
\partial_t
f^\e_i+\frac{1}{\varepsilon}v\cdot\nabla_xf^\e_i+\frac{1}{\varepsilon}
F^\e_i\cdot\nabla_v f^\e_i=\frac{1}{\varepsilon^2}L_{\beta}f^\e_i
\end{equation}
Here $ F^\e_i$ is the rescaled Vlasov term (\ref{0.3}) with $\gamma=\e$:
\[
 F^\e_i(x,t)=-\nabla_x\int_{\TT^d}{\rm d }x'
U(|x-x'|) \int_{\mathbb{R}^d}{\rm d }v f^\e_j(x',v,t) :=-\nabla_x U\star \rho^\e_j
\]
We substitute in (\ref{eq-res}) the formal power series  for
$f^\e_i$ and
$ F^\e_i$
\begin{eqnarray*}
&\displaystyle{f^\e_i=\sum_{n=0}^{\infty}\varepsilon^n
 {f}_i^{(n)},\qquad
  F^\e_i=\sum_{n=0}^{\infty}\varepsilon^n  F_i^{(n)}}&\\
 &\displaystyle{F_i^{(n)}=-\nabla_x U\star
 \int_{\mathbb{R}^d}}{\rm d }v  f_j^{(n)}(x',v,t)&
\end{eqnarray*}
We get
\begin{eqnarray*}
 &\displaystyle{\frac{1}{\varepsilon^2}L_{\beta}f_i^{(0)}+\frac{1}{\varepsilon}\left\{
 L_{\beta}f_i^{(1)}-v\cdot\nabla_xf_i^{(0)}-
 F_i^{(0)}\cdot\nabla_vf_i^{(0)}\right\}-}&\\
 &\displaystyle{\sum_{n=0}^{\infty}\varepsilon^n\left\{ \partial_t
 f_i^{(n)}+v\cdot\nabla_xf_i^{(n+1)}+\sum_{l,l'\ge
 0 :l+l'-1=n}\left(
 F_i^{(l)}\cdot\nabla_vf_i^{(l')}\right)-L_{\beta}f_i^{(n+2)}\right\}=0}&
\end{eqnarray*}
At each order in $\e$ we get an equation.   We write down here explicitly the first three orders:
\begin{enumerate}
 \item[$\varepsilon^{-2}$)] $L_{\beta}f_i^{(0)}=0$
 \item[$\varepsilon^{-1}$)]
 $v\cdot\nabla_xf_i^{(0)}+ F_i^{(0)}\cdot
 \nabla_vf_i^{(0)}=L_{\beta}f_i^{(1)}$
 \item[$\varepsilon^{0}$)]
 $\partial_tf_i^{(0)}+v\cdot\nabla_xf_i^{(1)}+
 F_i^{(0)}\cdot\nabla_vf_i^{(1)}+
 F_i^{(1)}\cdot\nabla_vf_i^{(0)}=L_{\beta}f_i^{(2)}$
\end{enumerate}
From $\varepsilon^{-2})$ we deduce from the properties of $L_\beta$ that $f_i^{(0)}$ is
the Maxwellian $M_{\beta}$ multiplied by a density factor depending on $x$ and $t$:
\begin{equation}\label{sol}
f_i^{(0)}=\rho_i(x,t) M_{\beta}
\end{equation}
Replacing this expression in the second equation (order  $\varepsilon^{-1}$) we get
\[
M_{\beta}v\cdot(\nabla_x\rho_i+\beta\rho_i\nabla_x(U\star\rho_j))=L_{\beta}f_i^{(1)}
\]
So a solution has to be of the form:
\[
f_i^{(1)}=M_{\beta}(A+B\cdot v)
\]
where $A$ will be fixed by the equations of the next orders
and $B$ is the vector
\[
B=-\frac{1}{\beta}(\nabla_x\rho_i+\beta\rho_i\nabla_x(U\star
\rho_j))
\]
If we put these expressions of $f_i^{(0)}$ and
$f_i^{(1)}$ in the $\varepsilon^0)$ equation and
integrate over $v$, remembering that
\[
\int_{\mathbb{R}^3}dvM_{\beta}
v_iv_j=\frac{1}{\beta}\delta_{ij},
\]
we find the equations for the zero order densities
\begin{equation}\label{density}
\partial_t
\rho_i-\frac{1}{\beta^2}(\Delta_x\rho_i+\beta\nabla_x\cdot(\rho_i\nabla_x(U\star
\rho_j)))=0.
\end{equation}

Now our aim is to show that a solution of equation (\ref{eq-res})
does exist and its limit as $\e$ goes to zero is
given by (\ref{sol}), with $\rho_i$ satisfying (\ref{density}). We try to solve (\ref{eq-res}) in terms of
a truncated  expansion
\begin{equation}\label{rem}
f^\e_i=\sum_{n=0}^K\varepsilon^n
f_i^{(n)}+\varepsilon^mR_i
\end{equation}

Replacing  expression (\ref{rem}) for $f^\e_i$ in
equation (\ref{eq-res}) we get:

For any $n$ between $0$ and $K$
\begin{eqnarray}
 &\displaystyle{\partial_tf_i^{(n-2)}+v\cdot\nabla_xf_i^{(n-1)}+}&\nonumber\\
 &\displaystyle{\sum_{l,l'\geq 0: l+l'-1=n-2}\left[-\nabla_x
  U\star \int_{\mathbb{R}^3}
 dvf_j^{(l)}\right]\cdot\nabla_vf_i^{(l')}-L_{\beta}f_i^{(n)}=0}&
 \label{exp}
\end{eqnarray}
$f_i^{(s)}=0, s\le 0$ and for the remainder
\begin{equation}\label{eq-rem}
\partial_{t} R_i +\frac{1}{\varepsilon}[v\cdot\nabla_x R_i+ F_i^{\e}\cdot\nabla_v R_i+
B_i\cdot\Gamma_i]=\frac{1}{\varepsilon^2}L_{\beta} R_i-\varepsilon^{K-m-1}A_i
\end{equation}
where we defined
\begin{eqnarray*}
 &\displaystyle{B_i=\sum_{n=0}^K\varepsilon^n\nabla_vf_i^{(n)},\qquad
\Gamma_i=-\nabla_x U\star
 \int_{\mathbb{R}^3}dv R_j}&\\
 &\displaystyle{A_i=\partial_tf_i^{(K-1)}+v\cdot
 \nabla_xf_i^{(K)}+\varepsilon\partial_t
 f_i^{(K)}+\sum_{n=K-1}^{2K-1} \varepsilon^{n-K+1}
 \sum_{0\leq l,l'\leq K\atop l+l'-1=n} F_i^{(l)}\cdot\nabla_v
 f_i^{(l')}}&\\
\end{eqnarray*}
We will find solutions
 $f_i^{(n)}$
 to equations  (\ref{exp}) in next section and  we will study the equation for the
remainder
$R_i$ in section 5. Here we state the results:

Denote by $(\cdot,\cdot)_-$ the following $L_2$ scalar product
 and with $||\cdot||_-$ the associated
norm
$$(h,g)_-:=\int_{\TT^d\times\mathbb R^d} dx dv M^{-1}_\beta(v)\sum_{i=1,2}[h_i(x,v)g_i(x,v)].$$ Put
$A=\{A_i\}_{i=1,2}$ and
$R=\{R_i\}_{i=1,2}$. Then
\begin{thm}
Given a classical solution $\rho_i(x,t)$ of the macroscopic equations (\ref{0.4}) in the time interval $[0,T]$,
there  is a constant
$C$ depending on $T$, such that a unique solution to (\ref{eq-rem}) exists and satisfies the bounds
\[
 \sup_{t\in[0,T]}||R(\cdot,t)||_-\le C\e^{K-1-m}||A||_-
\]
\label{resto}
\end{thm}
As a consequence,
\begin{coro}
Under the assumptions of Theorem \ref{resto}  and $m\ge 1, K-1-m\ge 0$ there is a positive constant $\e_0$
such that for
$\e<\e_0$ there is a smooth solution
$f_i^\e(x,v,t)$ to the rescaled Vlasov-Fokker-Plank equations (\ref{eq-res}) satisfying for some constant $C$
$$\sup_{t\in[0,T]}||f_i^\e- M_\beta\rho_i||_{ -}\le C \e$$
\end{coro}

\bigskip
\goodbreak

\section{Macroscopic limit: Expansion terms}
\setcounter{equation}{0}

In this section we show existence and regularity properties of $f_i^{(n)}$. For simplicity, we write down the
proof only for
$K=2$, but the argument goes on for any $K$. The structure of equations (\ref{exp}) is very simple:
they are of the form
\begin{equation}\label{eqn}
L_{\beta}f=h
\end{equation}
with $h$ a given function.   In the
Hilbert space with scalar product
$$(h,g)_M=\int_{\mathbb R^d} dv h(\cdot,v)g(\cdot,v)M_\beta^{-1}$$
the kernel  $\mathcal{N}=\ker(L_{\beta})$ is made of constants in velocity  multiplied by $M_{\beta}$.
  Hence  this equation has a
solution iff $h$ is in the orthogonal to the kernel of
$L_\beta$ namely iff
\begin{equation}
\int dv h(v)=0.
\label{comp}
\end{equation} Moreover, the solution is determined but for a term in the kernel
which is of the form a function of $x,t$ times the Maxwellian.   Starting from the lowest order, we
will see that
$h(x,v,t)= P(x,t,v)
 M_\beta$ with
$P$ a polynomial of the velocity with coefficients eventually depending on $ x,t$.
  $v$ The equation (\ref{eqn}) can be
solved uniquely in the orthogonal to the null space of $L_{\beta}$. If $P$ is a polynomial the
solution is again a polynomial of the same degree of $P$ multiplied by the maxwellian $M_{\beta}$. In
other words, if $M_{\beta}P\in
\mathcal{N}^{\bot}$ with $P$ a polynomial, then there exists a
unique $f\in \mathcal{N}^{\bot}$ such that (\ref{eqn}) holds. This
statement can be shown by finding explicitly solutions to the
problem (\ref{eqn}) for different choices of the polynomial $P$.
We are interested in polynomials of degree up to the second.

For $n=0,1$ the equations (\ref{exp}) are of the form (\ref{eqn}) with $h=0$ and $h=b_iv_i$
respectively and  have already been discussed in the previous section. We recall that    $f^{(1)}$
can be found as
$M_{\beta}(A_i +B_iv_i)$, with
$B_i=-\frac{1}{\beta}b_i$. $A_i$ would be   determined by the compatibility condition at the
order $n=3$. Since we are truncating the expansion at $n=2$ we can safely choose $A_i$ equal to zero.

Let us now deal with a polynomial of degree two:
\[
P(v)=a+b_iv_i+c_{ij}v_iv_j.
\]
By gaussian integration, the condition (\ref{comp}) becomes
\begin{equation}\label{trace}
a+\frac{1}{\beta}c_{ii}=0.
\end{equation}
We look for a solution of (\ref{eqn}) of the following type:
\[
f(v)=M_{\beta}(A+B_iv_i+C_{ij}v_iv_j).
\]
Plugging this ansatz in our equation we find
\[
\partial_{v_k}(M_{\beta}\partial_{v_k}(A+B_iv_i+C_{ij}v_iv_j))=M_{\beta}P(v).
\]
Recall that $\partial_{v_k}v_i=\delta_{ki}$ and
$\partial_{v_k}v_iv_j=\delta_{ki}v_j+\delta_{kj}v_i$; then the
left hand side of the above equation simplifies to
\begin{eqnarray*}
 &\displaystyle{\partial_{v_k}(M_{\beta}(B_k+C_{kj}v_j+C_{ik}v_i))=}&\\
 &\displaystyle{M_{\beta}(-\beta
 v_k(B_k+C_{kj}v_j+C_{ik}v_i)+\delta_{kj}C_{kj}+\delta_{ik}C_{ik})}=&\\
 &\displaystyle{M_{\beta}(2C_{ii}-\beta B_iv_i-2\beta
 C_{ij}v_iv_j)}&
\end{eqnarray*}
and identifying the coefficients of the corresponding powers of
$v_i$ one gets
\[
B_i=-\frac{1}{\beta}b_i\;\;C_{ij}=-\frac{1}{2\beta}c_{ij}
\]
with the  relation $a=2C_{ii}$ which is automatically
verified, thanks to the compatibility condition (\ref{trace}). In
order to fix the parameter $A$ we impose the analog of
(\ref{trace}): $A+C_{ii}/\beta=0$,  namely as before we are choosing equal to zero the projection on
the null space of
$L_\beta$. Thus
\[
A=-\frac{a}{2\beta}.
\]
In the context of our problem the known term is always in the form of a polynomial multiplied by a maxwellian and the coefficients of the $v_i$ are functions of the position.
In the case where only first powers of $v$ appear, i.e. $P(v)=a_k^{(i)}v_k$, the $a_k^{(i)}$ are given by
\[
a^{(i)}= \nabla_x\rho_i+\beta\rho_i\nabla_x(U\star  \rho_j)
\]
here $i,j=1,2$ and $i\neq j$. When $P(v)=a^{(i)}+b^{(i)}_kv_k+c_{hk}^{(i)}v_hv_k$ the coefficients are the following:
\begin{eqnarray*}
 &a^{(i)}=\partial_t\rho_i+\nabla_x(U\star  \rho_j)\cdot\frac{1}{\beta}(\nabla_x\rho_i+\beta\rho_i\nabla_x(U\star \rho_j))&\\
 &b_k^{(i)}=0&\\
 &c_{hk}^{(i)}=-\frac{1}{\beta}\partial_{x_h}(\partial_{x_k}\rho_i+\beta\rho_i\partial_{x_k}(U\star \rho_j))-\partial_{x_h}(U\star \rho_j)(\partial_{x_k}\rho_i+\beta\rho_i\partial_{x_k}(U\star \rho_j))&
\end{eqnarray*}
Summing up we denote by $f_i^{(k)}$, $k=0,1,2$ the following functions of $v$ and $x$:
\begin{eqnarray*}
 &f_i^{(0)}=M_{\beta}\rho_i(x,t)&\\
 &f_i^{(1)}=-\frac{1}{\beta}M_{\beta}v\cdot(\nabla_x\rho_i+\beta\rho_i\nabla_x(U\star \rho_j))&\\
&f_i^{(2)}=-\frac{1}{2\beta}M_{\beta}\left[\partial_t\rho_i+\nabla_x(U\star \rho_j)\cdot\frac{1}{\beta}(\nabla_x\rho_i+\beta\rho_i\nabla_x(U\star \rho_j))\right.&\\
&\left.-v\cdot\frac{1}{\beta}\nabla_x(v\cdot(\nabla_x\rho_i+\beta\rho_i(\nabla_x(U\star \rho_j)))-v\cdot\nabla_x(U\star \rho_j)v\cdot(\nabla_x\rho_i+\beta\rho_i\nabla_x(U\star \rho_j)))\right]&
\end{eqnarray*}
where $\rho_i$ is solution of
\[
\partial_t\rho_i-\frac{1}{\beta^2}(\Delta_x\rho_i+\beta\nabla_x\cdot (\rho_i\nabla_x(U\star \rho_j)))=0.
\]
The known term $A_i$ appearing in the equation for the remainder becomes
\[
A_i=\partial_tf_i^{(1)}+v\cdot\nabla_xf_i^{(2)}+\e\partial_tf_i^{(2)}+\sum_{n=1}^3\e^{n-1}\sum_{0\le l,l'\le
2\;l+l'-1=n}F_i^{(l)}\cdot\nabla_vf_i^{(l')}
\]
where we recall that
\[
 F_i^{(n)}=-\nabla_x\int_{\e\Omega}dx'U(|x-x'|)\int_{\mathbb{R}^3}dv'f_j^{(n)}
\]
It is easy to show that the sum over $l,l'$ is given by
\[
-\nabla_x(U\star \rho_j)\cdot\nabla_vf_i^{(2)}
\]
indeed $F_i^{(1)}=0=F_i^{(2)}$ because the functions $f_i^{(1)}$ and $f_i^{(2)}$
belong both to
$\mathcal{N}^{\bot}$ and $F_i^{(0)}=-\nabla_x(U\star \rho_j)$.

In conclusion, the $f_i^{(n)}$ are always of the form $M_\beta$ times a polynomial in $v$ times a function of
$x,t$ which depends on the derivatives of $\rho_i(x,t)$ solution of the macroscopic equations. If we fix an
initial datum for (\ref{0.4}) in $ C^2(\TT^d)$ then the corresponding unique solution will be classical as
shown in section 5 and the $f^{(n)}=\{f^{(n)}\}_i$ as well $A_i$ will satisfy the regularity properties
$$||f^{(n)}||_-\le C,\quad ||A||_-\le C$$.

\bigskip
\goodbreak

\section{Macroscopic limit: Remainder}
\setcounter{equation}{0}

In this section we will find a solution to equation (\ref{eq-rem}), which is a weakly non linear equation if
$m\ge 1, K-1-m\ge 0$, by considering first the linear problem with the force term
$F_i^{\e}$ assumed given so that  general results will grant  the existence of this linear
problem in a suitable space. Then,  a
fixed point argument applies by using $\e$ as small parameter. From here on we will simplify notation by setting $M=M_{\beta}$.

Define $\tilde f=f/M$ and
\[
\tilde L_{\beta}\tilde f=\frac{1}{M}\nabla_v\cdot(M\nabla_v(\tilde
f)).
\]
Moreover, we introduce the Hilbert space associated to  the $L_2$ scalar product $(\cdot,\cdot)_M$
weighted by the maxwellian and with $||\cdot||_M$ the associated
norm. In this Hilbert space $\tilde L_{\beta}$ is self-adjoint and non positive:
\begin{eqnarray*}
 &\displaystyle{(g_1,\tilde L_{\beta}g_2)_M=(\tilde
 L_{\beta}g_1,g_2)_M}&\\
 &\displaystyle{(g,\tilde L_{\beta}g)_M=\int_{\TT^d\times\mathbb R^d} dxdv
Mg\frac{1}{M}\nabla_v\cdot(M\nabla_v(g))=-||\nabla_v g||_M^2.}&
\end{eqnarray*}
If we put $R_i=\psi_i M$, the equation for the remainder becomes
\begin{equation}
\partial_t\psi_i+\e^{-1}\left[v\cdot\nabla_x\psi_i+\frac{F_i^{\e}\cdot\nabla_v(M\psi_i)}{M}+\frac{B_i\cdot\Gamma_i}{M}\right]=\e^{-2}\tilde
L_{\beta}\psi_i-\e^{K-1-m}\frac{A_i}{M}.
\label{RR}
\end{equation}
In order to estimate $||\psi_i||_M$ one multiplies the above
equation by $M\psi_i$ and integrates over $x$ and $v$. So the
first term on the left hand side becomes
\[
\frac{1}{2}\partial_t ||\psi_i||_{M}^2
\]
while the gradient with respect to the position disappear because
of the periodic boundary conditions.

We  assume that  the force terms $F_i^{\e}$ are given functions that we will call $\hat
F_i$ and are such that
\[
||\hat F_i||_{\infty}\le\alpha_{\hat F};
\]
Hence
\[
\left|\int dxdv \psi_i\hat
F_i\cdot\nabla_v(M\psi_i)\right|=\left|\int dxdv \hat
F_i\cdot(M^{\frac{1}{2}}M^{\frac{1}{2}}\psi_i\nabla_v\psi_i)\right|\le
||\hat F_i||_{\infty}||\psi_i||_M||\nabla_v\psi_i||_M
\]
where we integrated by parts ($\hat F_i$ depends only on $x$) and
we used Schwartz inequality. Now the term with the convolution of
the remainder with the gradient of the potential is estimated in
the following way:
\begin{eqnarray*}
 &\ &\left|\int dxdv \psi_i(x,v)\nabla_v(\rho(x)M(v))\cdot\int
 dx'\nabla_xU(|x-x'|)\int dv'M(v')\psi_j(x',v')\right|\\
 &=&\left|\int dxdv \rho
 M^{\frac{1}{2}}(v)M^{\frac{1}{2}}(v)\nabla_v\psi_i(x,v)\cdot \int
 dx'dv' \nabla_xU M^{\frac{1}{2}}(v')M^{\frac{1}{2}}(v')\psi_j(x',v')\right|\\
 &\le&|\TT^d|\;\sup_{\TT^d}|\rho|\;
 \sup_{\TT^d}|\nabla_xU|\;||\psi_j||_M\;||\nabla_v
 \psi_i||_M.
\end{eqnarray*}
As before we first integrated by parts and then we applied
Schwartz inequality twice. Here it has been considered  only the
lowest order in $\e$ of the sum which constitutes $B_i$; the other
terms are treated similarly. The last estimate is the one for $A_i$:
\begin{eqnarray*}
 &\displaystyle{\int dxdv \psi_i A_i=\int dxdv
 M^{\frac{1}{2}}\psi_i\frac{A_i}{M_{\frac{1}{2}}}\le}&\\
 &\displaystyle{||\psi_i||_M\;||M^{-1}A_i||_M\le
 \frac{1}{2}(||\psi_i||_M^2+||M^{-1}A_i||_M^2).}&
\end{eqnarray*}
Summing up, we have
\begin{eqnarray*}
 &\displaystyle{\frac{1}{2}\partial_t ||\psi_i||^2_{M}
 \le-\e^{-2}||\nabla_v\psi_i||_M^2+(c_1+\e^{-1}c_2)||\psi_j||_M||\nabla_v\psi_i||_M+
 \alpha_{\hat F_i}\e^{-1}||\psi_i||_M||\nabla_v\psi_i||_M}&\\
 &\displaystyle{+\frac{\e^{K-1-m}}{2}(||\psi_i||_M^2+||M_{-1}A_i||_M^2).}&
\end{eqnarray*}
Note that $c_1$ contains powers of $\e$ greater than $\e^{-1}$.
Now one exploits the inequality
\begin{equation}\label{ineq}
-\frac{\e^{-2}x^2}{2}+(\sigma_1+\e^{-1}\sigma_2)xy\le \frac{(\e
\sigma_1+\sigma_2)^2}{2}y^2
\end{equation}
but first we need to introduce the norm
$||\psi||_M^2:=||\psi_1||_M^2+||\psi_2||_M^2$, so we have
\begin{eqnarray*}
 \frac{1}{2}\partial_t ||\psi||_M^2&\le&
 -\e^{-2}(||\nabla_v\psi_1||_M^2+||\nabla_v\psi_2||_M^2)
+(c_1+\e^{-1}c_2)\big(||\psi_2||_M||\nabla_v\psi_1||_M\\
 &+&||\psi_1||_M||\nabla_v\psi_2||_M)
+\alpha_{\hat
 F}\e^{-1}(||\psi_1||_M||\nabla_v\psi_1||_M+||\psi_2||_M||\nabla_v\psi_2||_M)\\
 &+&\frac{\e^{K-1-m}}{2}[(||\psi_1||_M^2+||\psi_2||_M^2)+(||M_{-1}A_1||_M^2+||M_{-1}A_2||_M^2)]\\
 &\le&\frac{(\e
 c_1+c_2)^2}{2}(||\psi_1||_M^2+||\psi_2||_M^2)+\frac{\alpha_{\hat
 F}^2}{2}(||\psi_1||_M^2+||\psi_2||_M^2)\\
 &+&\frac{\e^{K-1-m}}{2}[(||\psi_1||_M^2+||\psi_2||_M^2)+(||M_{-1}A_1||_M^2+||M_{-1}A_2||_M^2)]
\end{eqnarray*}
where we used the inequality (\ref{ineq}) in two different ways;
in fact we divided the negative term in two halves and then once
we chose $c_1=\sigma_1$ and $c_2=\sigma_2$ and once we put $c_1=0$
and $c_2=\alpha_{\hat F}$.

Multiplying by $2$ both  members one gets
\[
\partial_t ||\psi||_M^2\le \lambda ||\psi||_M^2+d
\]
where $\lambda=\lambda(\alpha_{\hat F})=\alpha_{\hat F}^2+(\e
c_1+c_2)^2+\e^{K-1-m}$ and
$d=\e^{K-1-m}(||M^{-1}A_1||_M^2+||M^{-1}A_2||_M^2)$.
Integrating over the time, by the Gronwall inequality:
\begin{eqnarray*}
 &\displaystyle{f(t)\le K(t)+\lambda\int_0^t d\tau f(\tau)\le K(T)
 +\lambda\int_0^t d\tau f(\tau)}&\\
 &\displaystyle{\Longrightarrow\;\;f(t)\le K(T)e^{\lambda t}\le
 K(T)e^{\lambda T}}&
\end{eqnarray*}
where $f=||\psi||_M^2$, $K(t)=\int_0^t d\tau d(\tau)$ is a non
decreasing function of time and we used the initial condition
$f(0)=0$.

Now consider the sequence of forces
\[
\hat F_i^{(k)}=-\nabla_xU\star \int_{\mathbb{R}^3}{\rm d}v
\sum_{n=0}^K\e^nf_j^{(n)}-\e^m\nabla_xU\star \int_{\mathbb{R}^3}{\rm d} vR_j^{(k-1)}
\]
with $k\ge 1$ and $R_i^{(0)}=0$. Let $\alpha_k=\max\{||\hat
F_1^{(k)}||_{\infty},||\hat F_2^{(k)}||_{\infty}\}$, then
\[
\alpha_k\le \bar \alpha +\e^m C\int {\rm d}x{\rm d}v
|R_j^{(k-1)}|
\]
where $j$ is chosen such that it corresponds to the maximum in the
definition of $\alpha_k$ and
\[
\bar\alpha=\sup_{x\in\TT^d}\left|\nabla_xU\star \int {\rm d}v\sum_{n=0}^K\e^nf_j^{(n)}\right|.
\]
Write $\int dv |R_j^{(k-1)}|=\int {\rm d}v |M\psi_j^{(k-1)}|=\int {\rm d}v
M^{\frac{1}{2}}|M^{\frac{1}{2}}\psi_j^{(k-1)}|$; using Schwartz
inequality we get
\[
\int {\rm d}x
{\rm d}v|R_j^{(k-1)}|\le|\TT^d|^{\frac{1}{2}}||\psi_j^{(k-1)}||_M.
\]
Thus, recalling the estimate for $||\psi||_M^2$, we can conclude
that
\[
\alpha_k\le \bar\alpha+\e^m\mu(\alpha_{k-1})
\]
where the non decreasing function $\mu$ is defined by
$\mu(\alpha_{k})=C(|\TT^d|K(T)\exp(\lambda(\alpha_{k})
T))^{1/2}$. By induction on $k$ we show that $\alpha_k\le 2\bar\alpha$ $\forall k$. In fact
\[
\alpha_1\le\bar\alpha\le 2\bar\alpha.
\]
Then suppose $\alpha_{k-1}\le 2\bar\alpha$; we have
\[
\alpha_k\le\bar\alpha+\e^m\mu(\alpha_{k-1})\le
\bar\alpha+\e^m\mu(2\bar\alpha)\le 2\bar\alpha
\]
because we applied the inductive hypothesis, exploited the
monotonicity of $\mu$ and chose $\e$ so small that
$\e^m\mu(2\bar\alpha)\le\bar\alpha$.

Denote with $\delta\psi_i^{(k)}$ the difference
$\psi_i^{(k)}-\psi_i^{(k-1)}$. The equation solved by
$\delta\psi_i^{(k)}$ is
\begin{eqnarray*}
 &\displaystyle{\partial_t(\delta\psi_i^{(k)})+\e^{-1}\left[v\cdot\nabla_x
 (\delta\psi_i^{(k)})+\frac{\hat
 F_i^{(k)}\cdot\nabla_v(M\psi_i^{(k)})-\hat F_i^{(k-1)}\cdot
 \nabla_v(M\psi_i^{(k-1)})}{M}+\frac{B_i\cdot\Gamma_i}{M}\right]}&\\
 &\displaystyle{=\e^{-2}\tilde L_{\beta}(\delta\psi_i^{(k)})}&
\end{eqnarray*}
where is understood that $\Gamma_i$ contains $\delta\psi_j^{(k)}$
and no more $\psi_j$. Summing and subtracting the quantity $\hat
F_i^{(k)}\cdot\nabla_v(M\psi_i^{(k-1)})$ one has
\[
\hat
 F_i^{(k)}\cdot\nabla_v(M\psi_i^{(k)})-\hat F_i^{(k-1)}\cdot
 \nabla_v(M\psi_i^{(k-1)})=\hat
 F_i^{(k)}\cdot\nabla_v(M\delta\psi_i^{(k)})+\delta\hat
 F_i^{(k)}\cdot\nabla_v(M\psi_i^{(k-1)})
\]
where
\[
\delta\hat F_i^{(k)}=\hat F_i^{(k)}-\hat
F_i^{(k-1)}=-\e^m\nabla_xU\star \int
{\rm d}v'M\delta\psi_j^{(k-1)}.
\]
If one multiplies the equation for $\delta\psi_i^{(k)}$ by
$M\delta\psi_i^{(k)}$ and integrates in space and velocities, it
is possible to replicate the above estimates for the norm of the
remainder. Only one thing is worth noting: the known term with
$A_i$ is now replaced by the following quantity
\[
\int {\rm d}x{\rm d}v \delta\psi_i^{(k)} \delta\hat
F_i^{(k)}\cdot\nabla_v(M\psi_i^{(k-1)})=-\int {\rm d}x{\rm d}v
M\psi_i^{(k-1)}\delta\hat
F_i^{(k)}\cdot\nabla_v(\delta\psi_i^{(k)})
\]
which one estimates in this way:
\begin{eqnarray*}
 &\displaystyle{\e^m\left|\int {\rm d}x{\rm d}v
 M(v)(\psi_i^{(k-1)}\nabla_v\delta\psi_i^{(k)})(x,v)\int
 dx'\nabla_xU(|x-x'|)\int {\rm d}v'
 M(v')\delta\psi_j^{(k-1)}(x',v')\right|}&\\
 &\displaystyle{\le\e^m\sup|\nabla_xU|\left(\int
 {\rm d}x{\rm d}v|M\psi_i^{(k-1)}\nabla_v\delta\psi_i^{(k)}|\right)\left(\int
 {\rm d}x'{\rm d}v'M|\delta\psi_j^{(k-1)}|\right)}&\\
 &\displaystyle{\le\e^m|\TT^d|^{\frac{1}{2}}\sup|\nabla_xU|\;
 ||\psi_i^{(k-1)}||_M||\nabla_v\delta\psi_i^{(k)}||_M
 ||\delta\psi_j^{(k-1)}||_M}&\\
 &\displaystyle{\le\e^m
 c||\nabla_v\delta\psi_i^{(k)}||_M||\delta\psi_j^{(k-1)}||_M\le
 \frac{\e^m
 c}{2}(||\nabla_v\delta\psi_i^{(k)}||_M^2+||\delta\psi_j^{(k-1)}||_M^2).}&
\end{eqnarray*}
In $c$ the bound for $||\psi_i^{(k-1)}||_M$ is also present. In
brief we have the following situation:
\[
f_k'\le Cf_k+\theta f_{k-1}
\]
for some $C$; $\theta$ depends on $\e$ and is small as we like  if $m\ge 1$,
and of course $f_k=||\delta\psi^{(k)}||_M^2$ with the same
notation as above. By integrating in time and using Gronwall
inequality we obtain
\[
f_k\le\int_0^T\theta f_{k-1}e^{CT}\le\int_0^T\theta
e^{CT}\int_0^T\theta e^{CT}f_{k-2}\le...\le const(\theta
e^{CT}T)^k
\]
thus, by a standard argument, we conclude that the sequence
$\{\psi_i(k)\}$ is a Cauchy sequence and the limit $\psi$ is the unique solution of (\ref{RR}) with bounded
norm
 $||\psi||_M$.

\bigskip

\section{Limiting equation}
\setcounter{equation}{0}

We follow a strategy similar to the one used in the previous section: we consider first a linear
problem, prove existence for it and then use a fixed point argument to give the existence for the
full non linear equation. Since we do not have at our disposal a small parameter we use compactness
arguments and the Schauder Fixed Point Theorem \cite{E}.
  We  seek for weak solutions in the following sense:

Let $W$ be the Hilbert space 
$$
W(0,T;H^1,H^{-1}):=\{f:f\in L^2(0,T;H^1), \frac{df}{dt}\in L^2(0,T;H^{-1})\}.
$$
$H^1(\TT^d)$ and $H^{-1}(\TT^d)$ Sobolev spaces on the torus with norms $$|v|_2^2=\int_{\TT^d} |v|^2, \qquad
||v||_1^2= |v|_2^2+ |\nabla v|_2^2$$ $$||v||_{-1}=\sup_{u\in H^1}[2(u,v)- ||u||_1^2)]=\int dk\frac{|\hat
v(k)|^2}{1+k^2}$$ $(\cdot,\cdot)$ scalar product in $L^2$.
$$||v||_W^2=\int_0^T[||v(t)||_1^2+||v'(t)||_{-1}^2]dt$$ with $v'=dv/dt$. Let $W_1$ be the convex subset of $W$
$$W_1=\{v\in W: \int_{\TT^d} v(x,t)=1 \quad \hbox{ a.e in } [0,T]\}$$
 We say that $\rho$ is a weak solution  of the linear problem (\ref{linear}) below if for
$\bar\rho\in L^2(\TT^d)$ and for  all
$v\in H^1(\TT^d)$ and a.a.
$0\leq t\leq T$
$$\beta^2(v,\rho')+(\nabla  v,\nabla\rho +\beta\rho\nabla U\star   h )=0$$
 and $\rho(\cdot,0)=\bar\rho(\cdot)$.

We remark that since  $\rho\in W$ implies $\rho\in C([0,T];L^2(\TT^d))$ we have that $\rho(0)\in
L^2(\TT^d)$.
\begin{thm}
\label{exlin}
For any $h\in  L^1(\TT^d)$ and $\bar u\in L^2(\TT^d)$ there exists a unique
solution in
$W_1$ to the following Cauchy problem
\begin{eqnarray}
\beta^2 \partial_t u&=& \Delta u+\beta\nabla\cdot(u\nabla(U\star h))\cr
u(\cdot,0)&=&\bar u(\cdot)
\label{linear}
\end{eqnarray}
\end{thm}

{\it Proof}.\ Since $h\in L^1(\TT^d)$ and $\nabla U$ as well as $\nabla^2 U$ are bounded we have
$\nabla(U\star h)$ and $\nabla^2(U\star h)$ in $ L^\infty([0,T]\times
\TT^d)$. Hence by standard arguments \cite{E} there exists a solution in $W$. Since the
equation is in form of divergence,
the total mass is conserved so that the solution is in $W_1$.

Moreover, we have some useful a priori estimates for the solution of
\ref{linear} (indeed the proof of existence can be achieved by approximation methods and these a priori
estimates). Denote by
$|u|_2$ the norm in
$L^2(\TT^d)$ :$|u|_2^2=\int_{\TT^d}dx |u|^2(x,t)$. We have that
\begin{equation}
\frac{1}{2}\frac{d}{dt}|u|_2^2=- \frac{1}{\beta^2}|\nabla u|^2_2 - \frac{1}{\beta}\int_{\TT^d}{\rm d}x
u(x,t)\nabla u(x,t)\nabla(U\star h)(x,t)
\end{equation}
 Since $h\in L^1(\TT^d)$ and $\nabla U$ is bounded
$$\sup_{x,t}|\nabla(U\star
h)(x,t)|\le \bar c $$

Then, for any $\d>0$
\goodbreak
\begin{eqnarray}
\frac{1}{2}\frac{d}{dt}|u|_2^2&\le&- \frac{1}{\beta^2}|\nabla u|^2_2 + \frac{\bar c}{\beta} |\nabla
u|_2 | u|_2\cr &\le &- (1-\d)\frac{1}{\beta^2}|\nabla u|^2_2 +\frac{1}{4\d}{\bar c}|
u|_2^2
\end{eqnarray}
By Gronwall there exists a constant $C$ such that
$$ |u|_2^2\le |\bar u|_2^2 e^{Ct}$$
 so that
$$\int_0^T{\rm d}t|u(t)|_2^2\le C |\bar u|_2^2
,\quad \int_0^T|\nabla u|^2_2\le C |\bar u|_2^2$$
for some constant $C$. Here and below $C$ denotes a running constant.
Moreover,
$$||u'||_{-1}=\sup_{v\in H^1:||v||_1=1}\{-\int_{\TT^d}\nabla v[\frac{1}{\beta}
\nabla U\star
h)+\frac{1}{\beta^2}\nabla u\}\le \frac{\bar c}{\beta} |u|_2+\frac{1}{\beta^2}|\nabla u|_2$$
Hence
$$\int_0^T{\rm d}t||u'(t)||_{-1}^2\le  C|\bar u|_2^2$$

Consider now functions $u: \TT^d\to R^2$. We define the Hilbert space $W$ in this case as before, simply
using as scalar product $(\cdot,\cdot)$ the scalar product in $L^2(\TT^d; R^2)$. We use the same
notation for $W$ and $W_1$.  We say that $\rho=(\rho_1,\rho_2)$ is a weak solution  of
(\ref{0.4}) if for all
$v\in H^1(\TT^d; R^2)$ and a.e.
$0\leq t\leq T$
$$\beta^2(v,\rho_i')+(\nabla v,\nabla\rho_i +\beta\rho_i\nabla U\star\rho_j)=0$$
and $\rho(\cdot,0)=\rho(\cdot)$.

 Theorem\ref{exlin} defines a map $A$ from $L^2(0,T;L^2(\TT^d; R^2))$ in
itself by applying it to a set of two equations for $u_i, i=1,2$ with a given term
depending on $g_i,i=1,2$ in the following way
\goodbreak
\begin{eqnarray}
\beta^2 \partial_t u_i&=& \Delta u_i+\beta\nabla(u_i\nabla(U\star g_j))\cr
u_i(\cdot,0)&=&\bar u_i(\cdot)\cr
 i,j=1,2,&&
i\neq j
\end{eqnarray}
We use $g=(g_1,g_2)$ and $u=(u_1,u_2)$, $|g|_2^2=\sum_{i=1,2}|g_i|^2_2$. Then, since the $L^1$ norm of
$g_j$ is bounded by a constant times the  $L^2$ norm, namely $L^1([0,T],\TT^d) \in L^2([0,T],\TT^d)$,  there
exists a solution $u$  in
$W$ and we can write
$$A(g)=u$$
$$||A(g)||_W^2\le C|\bar u|_2^2$$

We now prove the existence theorem for the non linear set of equations by proving that $A$ is
continuous and maps  a closed convex set in a compact set.

\ {\it Compactness.} We consider the closed and convex set $X\in  L^2([0,T],\TT^d)$
$$X=\{h: ||h||^2_{L^2(0,T;L^2)}\leq k\}$$
Since $A(h)$ is in $W$ and $W$ is compactly imbedded
in
$L^2([0,T],\TT^d)$ the image of $X$ is compact.

{\it Continuity.}  Consider $g,\tilde g\in L^2(0,T;L^2)$. Let $u=A(g)$ and $\tilde u=A(\tilde g)$ the
corresponding weak solutions. We have that, for $i\neq j$
\begin{eqnarray}
(u_i-\tilde u_i,u_i'-\tilde u_i')=&-&\frac{1}{\beta^2}\int_{\TT^d}|\nabla (u_i-\tilde u_i)|^2
-\frac{1}{\beta}\int_{\TT^d}(u_i-\tilde u_i)\nabla (u_i-\tilde u_i)\cdot \nabla(U\star  g_j)
\cr&-&\frac{1}{\beta}\int_{\TT^d}\tilde u_i\nabla(u_i-\tilde u_i)\cdot\nabla U\star(g_j-\tilde g_j)
\end{eqnarray}

$$\frac{1}{2}\frac{d}{dt}|u_i-\tilde u_i|_2^2\leq -C|\nabla (u_i-\tilde u_i)|^2_2 +c_1|u_i-\tilde
u_i|^2_2 + c_2|g_j-\tilde g_j|^2_2 $$
We have used that the $L^1$ norm of $(g-\tilde g)$ is bounded by the $L^2$ norm. Therefore,
$$||u-\tilde u||_{L^2([0,T],L^2)}\le C ||g-\tilde g||_{L^2([0,T],L^2)}$$
which proves the continuity of $A$ in ${L^2([0,T],L^2)}$.

By  Schauder's theorem the map $A$ has a fixed point in ${L^2([0,T],L^2)}$ which is the weak solution
we were looking for.

{\it Uniqueness} The proof is standard [\cite{GL}].

Summarizing, we have proved the following
\begin{thm}
\label{exist}
There exists a unique weak solution  in $W_1$ to the following Cauchy problem
\begin{eqnarray}
\beta^2 \partial_t \rho_i&=& \Delta\rho_i+\beta\nabla(\rho_i\nabla(U\star\rho_j)),\cr
\rho_i(\cdot,0)&=&\bar\rho_i(\cdot)\cr
i,j=1,2,&&
i\neq j
\end{eqnarray}
\end{thm}

{\it Regularity.} If $\nabla U\star \rho\in C^0([0,T];C^1)$ and $\bar \rho\in C^2(\TT^d)$ then the linear
equation has a classical solution. Since the weak solution $\rho$ is also in $C^0([0,T];L^2)$ we have
that indeed $\nabla U\star \rho\in C^0([0,T];C^1)$ and therefore the weak solution $\rho$ corresponding
to an initial datum in $C^2(\TT^d)$ is a classical solution.

\bigskip

\section{ Sharp interface limit}
\setcounter{equation}{0}

In this section we study the solutions of (\ref{0.2})  in the sharp interface limit in a  $3$-d torus
$\Omega$.
 We introduce again   the scale separation parameter
$\e$, which has the meaning of ratio between the kinetic and macroscopic scales. Then, we scale position and time
as $\e^{-1}$ and $\e^{-3}, $ respectively, while keeping fixed (equal to $1$) $\gamma$.  The width of the interface
on the macroscopic scale is then of order $\e$, so that in the limit $\e\to 0$ the interface becomes sharp.   The
rescaled density distributions
$
f_i^{\e}(r,v,t)=f_i(\e^{-1}r,v,\e^{-3}t) $, are solutions of
\begin{equation}\label{maineq}
\partial_{t}f_i^{\e}+\e^{-2} v\cdot\nabla_r f_i^{\e}+\e^{-2} F_i^{\e}\cdot\nabla_v
f_i^{\e}=\e^{-3}L_{\beta}f_i^{\e}.
\end{equation}
\[
F_i^{\e}(r,t)=-\nabla_r\int {\rm d}r'\e^{-3}U(\e^{-1}|r-r'|)\int {\rm d}v'f_j^{\e}(r',v',t)=:-\nabla_r g_i^{\e}.
\]
In this section $F_i^{\e}$ depends on $\e$ through the function $f_j^{\e}$ but also through the potential since we
are keeping fixed $\gamma$. We consider  a situation in which initially an interface is present. Since the
stationary non homogeneous solutions of  (\ref{0.2})
 are given by the Maxwellian multiplied by the front density profiles  we let our system start initially
close to those stationary solutions and   choose as initial datum  $f_i^\e(r,v)=M_\beta(v)\rho_i^\e$,
 where the density profiles are
very close to a profile such that in the bulk its values are $\rho_i^{\pm}$, the values of the densities in the two
pure phases at temperature $T$, and the interpolation between them on the interface is realized along the normal
direction in each point by the fronts. We put $\rho_1^{\pm}=\bar\rho^{\pm}$ and use the symmetry properties of the
segregation phase transition giving $\rho_2^{\mp}=\bar\rho^{\pm}$. Consider a smooth surface $\Gamma_0\subset
\Omega$.  Let $d(r,\Gamma_0)$ be the signed distance of the  point $r\in \Omega$  from the interface. Consider an
initial profile for the densities $\rho_i^\e$ of the following type: at distance greater than $O(\e)$ from the
interface
 (in the bulk) the density profiles $\rho^\e_i(r)$ are almost constant equal to $\rho_i^\pm$; at distance $O(\e)$
(near the interface) we choose
\begin{equation}\rho^\e_i(r)= w_i(\e^{-1}d(r,\Gamma_0))+O(\e)
\label{datum}
\end{equation}
where $w_i(z)$ are the fronts, which are one dimensional
  solutions of (\ref{0.1})  with asymptotic values $\rho_i^\pm$. Since these
solutions are unique up to a translation we fix a solution by imposing that
 $w_1(0)=w_2(0)$.

 Let  $\Gamma^\e_t$ be an interface at time $t$ defined by
 $$\Gamma^\e_t=\{r\in\Omega: \rho^\e_1(r,t)=\rho^\e_2(r,t)\}$$
and $T$ be such that $\Gamma^\e_{t}$ is regular for $t\in[0,T]$. Let $d^\e(r,t)$ be  the signed distance $d
({r},\Gamma^\e_t)$ of ${r}\in\Omega $ from the interface $\Gamma^\e_t$, such that $d^\e >0$ in $\Omega^{\e,+}_t$
and $d^\e <0$ in $\Omega^{\e,-}_t$, where  $\Omega=\Gamma^\e_t\cup \Omega^{\e,+}_t\cup\Omega^{\e,-}_t$.  For sake
of simplicity we drop from now on the apex $\e$.
 For any $r$ such that $|d(r,t)|<\frac{1}{k(\Gamma_t)}$, $ k(\Gamma_t)=
\sup_{x\in\Gamma_t}k(x)$ with $k( x)$  the maximum of the principal curvatures in $ x$, there exists $
s(r)\in\Gamma_t$ such that $$\nu( s(r))d(r,t) +s(r)=r$$ where $\nu( s(r))$ is the normal to the surface $\Gamma_t$
in $s(r)$. Hence, $$\nu(s(r))=\nabla_r d(r,t), \quad r\in \Gamma_t.$$
  Define the normal velocity of the interface as
$$V(s(r))={\partial_t}d(r,t).$$
 The  curvature $K$ (the sum of the principal curvatures)  is given by $K=\Delta_r
d(r,t), \quad r\in \Gamma_t.$  Define, for $\e^0$ small enough, $$\mathcal {N}(\delta):=\{r: |d(r,t)|<\delta\}$$
where $\delta=\frac{1}{n},\quad n=\max_{t\in[0,T],0\le\e\le\e^0}k(\Gamma_t)$.

We follow the  approach  based on  the truncated Hilbert expansions introduced by Caflish \cite{C}. This method,
which  has been used  in the previous chapter to prove the hydrodynamic limit for  the Vlasov-Fokker-Planck
equation, has been improved by including boundary layer expansions in [ELM], to  prove the hydrodynamic limit for
the Boltzmann equation in a slab.  Here we try to adapt the arguments in [ELM] to the fact that the boundary is not
given a priori and has to be found as a result of the expansion. The Hilbert expansion is nothing but  a power
expansion in $\e$ for the solution of the kinetic equation
\begin{equation}f^\e=\sum_{n=0}^\infty\e^nf^{(n)} .
\label{expansion}
\end{equation}
Since we expect that the behavior of the solution will be different in the bulk and near the interface, we
decompose $f^{(n)}$ in two parts: the bulk part $\hat f^{(n)}(r,t)$  and boundary terms  $\tilde f^{(n)}$ which
will be fast varying functions close to the interface, namely they depend  on $r,t$  in the following way $$\tilde
f^{(n)}=\tilde f^{(n)}(\e^{-1}d(r,t),r,t)$$
 while  $\hat f^{(n)}(r,t)$ are slowly varying
functions  on the microscopic scale. More precisely, a fast varying function $h(r,t)$ for $r\in {\mathcal N}$ can
be represented as a function $h(z,r,t)$, $z=\e^{-1} d(r,t)$, with the condition $h(z,r+\ell \nu(s(r)),t)=h(z,r,t),
\forall \ell$ small enough. Hence in $\mathcal {N}$
 we can write
\begin{equation}
 \nabla_r h=\frac{1}{\e}\nu\partial_z h+ \overline{\nabla}_r h; \quad
 \partial_{t}h=\frac{1}{\e}V\partial_z h + \partial_t h;\quad
\Delta_r h=\frac{1}{\e^2}\partial_z^2 h+\frac{1}{\e}(\nabla_r\cdot\nu)\partial_z h+\overline{\Delta}_r h
\label{above}
\end{equation}
where the bar on the derivative operators means  derivatives with respect to $r$, keeping fixed the other
variables. Note that $\nu\cdot \overline{\nabla}_r h(z,r,t)=0$.

 To write the expansion for the force term  $F_i^\e$ we
introduce  $U^\e\star \sum_{n=0}^\infty\e^n\rho_j^{(n)}=\sum_{n=0}^\infty\e^ng_i^{(n)} $ and $F_i^{(n)}=-\nabla_r
g_i^{(n)}$.
 We expand also the signed distance
\begin{equation}
d(r,t)=\sum_{i=0}^{\infty}\e^nd^{(n)}(r,t) \label{N}
\end{equation}
We will denote by $\nu^{(n)}$ the gradient $\nabla_r d^{(n)}$, with $\bar\nu:=\nu^{(0)}$.
 The condition $|\nabla_r d |^2=1$ is equivalent to: $$|\nabla_r d^{(0)}|^2=1,\quad \nabla_r d^{(0)}\nabla_r
d^{(1)}=0,\quad \nabla_r d^{(0)}\nabla_r d^{(j)}=-\frac{1}{2}\sum_{i=1}^{j-1}\nabla_r d^{(i)}\nabla_r d^{(j-1)},
\quad j\geq 2$$
 so that
$d^{(0)}$ can be interpreted as a signed distance from an interface that we denote by   $\bar\Gamma_t$.  As a
consequence of (\ref{N}) the velocity of the interface $\Gamma_t$ has the form $$\sum_{i=0}^{\infty}\e^i
V^{(i)},\quad \bar V:=V^{(0)}.$$
We remark that giving the velocity $V$  determines the curve evolving with it. The velocity $\bar V$ will generate
an order zero interface $\bar\Gamma_t$. The interface generated by $\sum_{i}\e^iV^{(i)}$ will be a deformation,
small for small $\e$, of $\bar\Gamma_t$.  We define $$\mathcal {N}^0(m):=\{r: |d^{(0)}(r,t)|<m \},
\bar\Gamma_t:=\{r: |d^{(0)}(r,t)|=0 \}, \Omega^{+(-)}:=\{r: |d^{(0)}(r,t)|> ( <) 0 \}$$ and fix $m$ so that
$\mathcal {N}^0(m)\subset \mathcal {N}(\delta)$.

We assume that in $\Omega^\pm\setminus \mathcal N^0({m})$
\begin{equation}f^\e=\sum_{n=0}^\infty\e^n \hat f^{(n)} .
\label{expansion0}
\end{equation}
and  that
 in $\mathcal { N}^0(m)$,  the solution is of the form
 \begin{equation}f^\e=\sum_{n=0}^\infty\e^n \tilde f^{(n)}
\label{expansion1}
\end{equation}
We will match the inner and outer expansions in $z=\e^{-1}m$ with $m=\e^{c}, c\in (0,1)$. Hence,
we require that as $z\to\pm\infty$ \cite{CF}
\begin{eqnarray*}
 &\displaystyle{\tilde f_i^{(0)}=(\hat
f_i^{(0)})^\pm+O(e^{-\a|z|})}\ \ \qquad\qquad\qquad\qquad\qquad\qquad\qquad\qquad\qquad
\qquad\qquad\quad\qquad\qquad\quad\ &\\
 &\displaystyle{  \tilde f_i^{(1)}=(\hat f_i^{(1)})^\pm+\nu^{(0)}\cdot(\nabla_r\hat
 f_i^{(0)})^\pm(z-d^{(1)})+O(e^{-\a|z|})}\ \  \qquad\qquad\qquad\qquad\qquad\qquad\qquad\qquad\quad&\\
 &\displaystyle{\tilde f_i^{(2)}=(\hat f_i^{(2)})^\pm+\nu^{(0)}\cdot(\nabla_r\hat
 f_i^{(1)})^\pm(z-d^{(1)})
+(\nabla_r\hat
 f_i^{(0)})^\pm\cdot(-\nu^{(0)}d^{(2)}+\nu^{(1)}(z-d^{(1)}))}\qquad\qquad\quad\ &\\
 &\displaystyle{+\frac{1}{2}(\partial_{r_h}
 \partial_{r_k}\hat
 f_i^{(0)})^\pm\nu^{(0)}_h(z-d^{(1)})\nu^{(0)}_k(z-d^{(1)})+O(e^{-\a|z|})}&\\
 &\displaystyle{\tilde f_i^{(3)}=(\hat f_i^{(3)})^\pm+\nu^{(0)}\cdot(\nabla_r\hat
 f_i^{(2)})^\pm(z-d^{(1)})+(\nabla_r\hat
 f_i^{(1)})^\pm\cdot(-\nu^{(0)}d^{(2)}+\nu^{(1)}(z-d^{(1)}))}\qquad\qquad&\\
 &\displaystyle{\quad+\frac{1}{2}(\partial_{r_h}\partial_{r_k}\hat
 f_i^{(1)})^\pm\nu^{(0)}_h\nu^{(0)}_k(z-d^{(1)})^2+(\nabla_r\hat
 f_i^{(0)})^\pm\cdot(\nu^{(2)}(z-d^{(1)})-\nu^{(0)}d^{(3)}-\nu^{(1)}d^{(2)})}&\\
 &\displaystyle{+\ (\partial_{r_h}\partial_{r_k}\hat
 f_i^{(0)})^\pm\nu^{(0)}_h(z-d^{(1)})(-\nu^{(0)}_kd^{(2)}+\nu^{(1)}_k(z-d^{(1)}))}&\\
 &\displaystyle{\!\!\!\!\!+\frac{1}{6}(\partial_{r_h}\partial_{r_k}\partial_{r_l}\hat
 f_i^{(0)})^\pm(\nu^{(0)}_h\nu^{(0)}_k\nu^{(0)}_l(z-d^{(1)})^3+O(e^{-\a|z|})}&\\
 &\ldots&
\end{eqnarray*}
where the symbol $(h)^\pm$ for the hat functions stands for $\lim_{\ell\to 0^\pm} h(r+\nu\ell)$, $r\in\bar\Gamma_t$ and the same for the derivatives.
  We replace
(\ref{expansion0}) and (\ref{expansion1}) in the equations and equate terms of the same order in $\e$ separately in
$\Omega^\pm\setminus \mathcal N^0(m)$ and $\mathcal { N}^0(m)$. We will use the  notation $\rho
_i^{(n)}=\int dv   f_i^{(n)}, $ and we denote by $\hat h,\tilde h$  a function $h (f_i^{(n)})$ whenever is
evaluated on $\hat f_i^{(n)},\tilde f_i^{(n)}$.

{\it Outer expansion}

\noindent In $\Omega^\pm\setminus \mathcal N^0(m)$, $n\ge 0$
\begin{equation}
 \displaystyle{\partial_t \hat f_i^{(n-3)}+v\cdot\nabla_r \hat f_i^{(n-1)}+
\sum_{l,l'\geq 0: l+l'=n-1}\hat F_i^{(l)}\cdot\nabla_v\hat f_i^{(l')}=L_{\beta}\hat f_i^{(n)}}, \label{bulk}
\end{equation}
with  $\hat f_i^\a=0, \a<0$.

{\it Inner expansion}

\noindent In $\mathcal { N}^0(m)$ $n\ge 0$ we have
\begin{eqnarray}
 \displaystyle{\sum_{l,l'\geq 0: l+l'=n-2}V^{(\ell')}\partial_z \tilde f_i^{(l)}+
\sum_{k+k'=n}\nu^{(k)}\cdot v\partial_z \tilde f_i^{(k')}+  v\cdot\overline{\nabla}_r  \tilde
f_i^{(n-1)}+\partial_t \tilde f_i^{(n-3)} }&\nonumber \\  \displaystyle{- \sum_{l,l',l''\geq 0:
l+l'+l''=n}\partial_z \tilde g_i^{(l)}\nu^{(l')}\cdot\nabla_v \tilde f_i^{(l'')} +\sum_{l,l'\geq 0:
l+l'=n-1}\overline{\nabla}_r \tilde g_i^{(l)}\cdot\nabla_v \tilde f_i^{(l')}=L_{\beta} \tilde f_i^{(n)}}\label{bb},
\label{boundary}
\end{eqnarray}
with  $\tilde f_i^\a=0, \a<0$.

The strategy for a rigorous proof
 is to construct, once the functions $ f_i^{(n)}$ have been determined, the solution in terms of a truncated Hilbert
expansion as
\begin{equation}f^\e=\sum_{n=0}^N\e^nf^{(n)}+\e^m R
\end{equation}
where the functions are evaluated in  $z=\e^{-1}d^N(r,t)$, with $ d^N(r,t)=\sum_{i=0}^{N-2}\e^n d^{(n)}(r,t)$ and
then write a weakly non linear  equation for the remainder. In this approach it  is essential to have enough
smoothness for the  terms of the expansion. On the contrary, they
 would be  discontinuous on the border of $\mathcal N^0(m)$ since $\tilde f^{(n)}$ are not exactly equal
to $\hat f ^{(n)}$ there but differ  for terms exponentially small in $\e$. One can modify the expansion terms by
interpolating in a smooth way between the outside and the inside getting smooth terms which do not satisfy the
equations for terms  exponentially small in $\e$, that can be put in the remainder.  With this in mind, we did not
put in the equations the terms coming from the force such that in the convolution $r$ is in $\mathcal N^0(m)$ and
$r'$ in  $\Omega^\pm\setminus\mathcal N^0(m)$. That is possible because the potential is of finite range. Finally, we   remark
that the terms $ f_i^{(n)}$ of the expansion do not depend on $\e$ but for being computed on $z$, which  depends on $\e$ because
of the rescaling and also because   the interface at time $t$ still depends  on $\e$. The latter is a new feature in the
framework of the Hilbert expansion  due to the fact that the boundary is not fixed  but is itself  unknown.

In this section we show how to construct the terms $f_i^{(n)}$. The argument is formal because we  do not
 prove boundedness of the remainder nor the regularity properties  of the terms of the expansion. We plan
to report on that in the future.

 Now we go back to the Hilbert power series  and start
examining the equations order by order. We will find explicitly only the first three terms in the expansion to
explain the procedure.

{\it Outer expansion}

 At the lowest order  $\e^{-3}$  ($n=0$):
$$L_{\beta} \hat f_i^{(0)}=0$$ which implies that $\hat f_i^{(0)}$ has to be Maxwellian in velocity with variance
$T$ times a function $\hat \rho_i^{(0)}(r,t)$. The latter is found by looking at the equations  at the next two
orders.
 At  order
$\e^{-2}$ ($n=1$):
\begin{equation}
 v\cdot\nabla_r \hat f_i^{(0)}+\hat F_i^{(0)}\cdot\nabla_v \hat f_i^{(0)}=L_{\beta}\hat f_i^{(1)}.
\end{equation}
The solution is of the form
\begin{equation}
\hat f_i^{(1)}=\hat\rho_i^{(1)}M_{\beta}-M_{\beta} \hat\rho_i^{(0)}v\cdot\nabla_r\hat\mu_i^{(0)} \label{f1}
\end{equation}
where $ \mu_i^\e(\rho^\e)=T\ln \rho_i^\e+U^\e\star \rho_j^\e$ and  $ \mu_i^\e=\sum_{n=0}^\infty\e^n\mu_i^{(n)}$ .

The order $\e^{-1}$ equation   $(n=2)$ is
 \begin{equation}
v\cdot\nabla_r \hat f_i^{(1)}+\hat F_i^{(0)}\cdot\nabla_v \hat f_i^{(1)}+\hat F_i^{(1)}\cdot\nabla_v \hat
f_i^{(0)}=L_{\beta} \hat f_i^{(2)} \label{f2}.
\end{equation}
 The solvability condition for this equation says that the integral on the velocity of the l.h.s.
has to be zero. By integrating  over the velocity and using the explicit expression for $\hat f_i^{(1)}$ we get
\[
-T\nabla_r\cdot(\hat \rho_i^{(0)}\nabla_r \hat\mu_i^{(0)})=0 .
\]
Hence the solvability condition for the equation $n=2$ gives the equation determining $\hat \rho_i^{(0)}$. The
choice of the initial data implies that the only solution of that equation is the constant one, with values
$\rho^\pm_i$ in $\Omega^{\pm}$.
  We look at next order  $n=3$   to find $\hat\rho_i^{(1)}$ by the
solvability condition.  By integrating over $v$  the equation $n=3$  and taking into account that $\hat f_i^{(0)}$
is Maxwellian in velocity, we get the following condition on $\hat u^{(2)}$,  where
 $u_i^{(n)}=\int dv  v f_i^{(n)}$,
\begin{equation}
\nabla_r \cdot \hat u_i^{(2)}=0\label{div}.
\end{equation}
Then,
 $\hat f_i^{(2)}$ is determined,  by replacing (\ref{f1}) in
 equation (\ref{f2}), as
\begin{equation}
\hat f_i^{(2)}=- M_{\beta}\hat\rho^{(0)}_iv\cdot\nabla_r\hat\mu^{(1)}_i +\hat \rho^{(2)}_i M_{\beta}. \label{f22}
\end{equation}
where  $\hat\mu_i^{(1)}=T\hat\rho_i^{(1)}/\hat\rho_i^{(0)}+\hat g_i^{(1)}$.

We use $\hat f_i^{(2)}$ as given by (\ref{f22}) to get  $\hat u_i^{2}=-T\hat \rho^{(0)}_i\nabla_r\hat\mu^{(1)}_i$
and plug it in (\ref{div}) to get the equation for $\hat \mu^{(1)}_i$ $$\Delta_r\hat\mu^{(1)}_i=0.$$ We consider
equation (\ref{bulk}) for $n=3$
\begin{equation}
 \displaystyle{\partial_t \hat f_i^{(0)}+v\cdot\nabla_r \hat f_i^{(2)}+
\sum_{l,l'\geq 0: l+l'=2}\hat F_i^{(l)}\cdot\nabla_v\hat f_i^{(l')}=L_{\beta}\hat f_i^{(3)}},
\end{equation}
whose solution is
\begin{equation}\hat f_i^{(3)}=-  M_{\beta}v\cdot[\hat\rho^{(0)}_i\nabla_r\hat\mu^{(2)}_i+
\hat\rho^{(1)}_i\nabla_r
\hat\mu^{(1)}_i]+M_{\beta}\frac{T}{2}\hat\rho^{(0)}_i(v\cdot\nabla_r)(v\cdot\nabla_r)\hat\mu^{(1)}_i+\hat
\rho^{(3)}_i M_{\beta} . \label{f3}
\end{equation}The equation for $\hat \mu^{(2)}_i=T\hat\rho_i^{(2)}/\hat\rho_i^{(0)}-
T/2(\hat\rho_i^{(1)}/\hat\rho_i^{(0)})^2+\hat g_i^{(2)}$  comes from
the equation  for $n=4$ $$
\partial_t\hat f_i^{(1)}+v\cdot\nabla_r\hat f_i^{(3)}-\sum_{l,l'\ge 0,\;l+l'=3}\nabla_r\hat
g_i^{(l)}\cdot\nabla_v\hat f_i^{(l')}=L_{\beta}\hat f_i^{(4)} $$ which gives as solvability condition
$\nabla_r\cdot \hat u_i^{(3)}=-\partial_t\hat\rho^{(1)}_i$ where $\hat u_i^{(3)}=\int dv \hat f_i^{(3)}$. By using
(\ref{f3}) we get $$\Delta
\hat\mu^{(2)}_i=\frac{1}{T\hat\rho_i^{(0)}}\partial_t\hat\rho^{(1)}_i-\frac{\nabla_r\hat\rho_i^{(1)}\cdot
\nabla_r\hat\mu_i^{(1)}}{\hat\rho_i^{(0)}}:=\frac{S_i}{\hat\rho_i^{(0)}},\;\;S_i=\beta\partial_t
\hat\rho_i^{(1)}-\nabla_r\hat\rho_i^{(1)}\cdot\nabla_r\hat\mu_i^{(1)} $$

{\it Inner expansion}

\noindent At the lowest order  ($n=0$) $$v\cdot\bar\nu\partial_z \tilde f_i^{(0)}-\bar\nu\cdot\nabla_v \tilde
f_i^{(0)}\partial_z \tilde g_i^{(0)} =L_{\beta} \tilde f_i^{(0)} .$$
 In  Appendix A it is proved that any
solution of this equation has  the form $ M_\beta(v)\tilde  \rho_i^{(0)}$, with $\tilde\rho_i^{(0)}$ a function of
$z$.
 Plugging back in the equation we have
\begin{equation}\label{istantone}
\partial_z \tilde\rho_i^{(0)}+\beta\tilde\rho_i^{(0)}\partial_z
(\tilde U\star \tilde\rho_j^{(0)})=0\;\;\Longleftrightarrow\;\;\partial_z\tilde\mu_i^{(0)}=0,
\end{equation}
  where $\tilde U$ is the potential $U$ integrated over all coordinates but one.  We solve
this equation with
 the conditions at infinity $\rho_i^{\pm}$, given by the matching conditions, and call $w_i$ this front solution.
  The  exponential decay of $w_i$ has been proved for the one-component case
\cite{DOPT} and  the same argument should provide the proof also in this case.
 We can conclude that in $\Omega$
$$f_i^{(0)}(r,t)=M_\beta[w(\frac{d(r,t)}{\e})\chi_{m}+(1-\chi_{m})\hat\rho_i^{(0)}],$$ with $\chi_{m}$ the
characteristic function of $\mathcal {N}^0(m)$. This solution differs from the front solution $w_i$ in $\Omega$ for
terms which are exponentially small in $\e$ and has the disadvantage of  not being continuous on the border of
$\mathcal N^0$.  As explained before, it has to
 be modified as
$$f_i^{(0)}(r,t)=M_\beta[w(\frac{d(r,t)}{\e})h(d(r,t))+(1-h(d(r,t)))\hat\rho_i^{(0)}(r,t)]$$ with $h$ a smooth
version of $\chi_m$.

We now find $\tilde f_i^{(1)}$  by examining the $\e^{-2}$ order
 (n=1)

\begin{equation}
v\cdot\bar\nu\partial_z \tilde f_i^{(1)}-\bar\nu\cdot\nabla_v \tilde f_i^{(0)}\partial_z\tilde g_i^{(1)}
-\bar\nu\cdot\nabla_v \tilde f_i^{(1)}\partial_z\tilde g_i^{(0)} =L_{\beta} \tilde f_i^{(1)} . \label{aab}
\end{equation}
The term involving $\nu^{(1)}$, $\nu^{(1)}\cdot (v\partial_z \tilde f_i^{(0)}-\nabla_v \tilde f_i^{(0)}\partial_z
\tilde g_i^{(0)})=\beta v\cdot\nu^{(1)}M_{\beta}\tilde\rho_i^{(0)}\partial_z\tilde\mu_i^{(0)}=0$, because $\tilde
f_i^{(0)}$ is solution of the  lowest order equation and the bar operators vanish because $\tilde\rho_i^{(0)}$ is
function of $z$ only. In Appendix A we show that the solution is necessarily  Maxwellian in velocity
so that we can write $\tilde f_i^{(1)}= \tilde\rho_i^{(1)}M_{\beta}$ with $\tilde \rho_i^{(1)}$ to be determined by
the following equation
\begin{equation}\label{e-2}
\partial_z\tilde\rho_i^{(1)}+\beta\tilde\rho_i^{(0)}\partial_z\tilde g_i^{(1)}+
\beta\tilde\rho_i^{(1)}\tilde U\star  \partial_z\tilde\rho_j^{(0)}=0 .
\end{equation}
  Taking into account that $-\beta\tilde U\star
\partial_z\tilde\rho_j^{(0)}=\partial_z\ln w_i$, from the equation for the front, we get
\begin{equation}
\partial_z\left(T\tilde\rho_i^{(1)}(w_i)^{-1}+\tilde
g_i^{(1)}\right)=0 \Longleftrightarrow\;\;\partial_z\tilde\mu_i^{(1)}=0 .
 \end{equation}
Hence, the value of $\tilde\mu_1^{(1)}-\tilde\mu_2^{(1)}$ in $z=0$  is enough to find
$\tilde\mu_1^{(1)}-\tilde\mu_2^{(1)}$  for any $z$. This value is found as follows. From
$$\tilde\mu_i^{(1)}=T(\tilde\rho_i^{(1)})(w_i)^{-1}+\tilde U\star \tilde \rho_j^{(1)}+{\bar K}\int  {\rm d}z' (z-z')\tilde
U(z-z')w_{j}(z'),$$
 where $\bar K=\Delta_r d^{(0)}(r,t)$ is the curvature of the interface $\bar \Gamma_t$ (see Appendix
 C), we want to find $\tilde\rho_i^{(1)}$ as determined by $\tilde\mu_i^{(1)}$. We define the operator
$\mathcal L$ as $(\mathcal L h)_i= T h_i (w_i)^{-1}+ \tilde U \star  h_j$. The previous relation reads as
\begin{equation}(\mathcal L \tilde\rho^{(1)})_i= \tilde\mu_i^{(1)}-{\bar K}\int  dz' (z-z')\tilde
U(z-z')w_{j}(z'). \label{ccc}
\end{equation}
 The  operator
$\mathcal L$ has a zero mode since $\mathcal L w'=0$, so that the equation $(\mathcal L\tilde\rho^{(1)})_i=h_i$ has
a solution only if $$\sum_{i=1,2}\int {\rm d}zh_i(z)w'_i(z)=0.$$ The solvability condition for (\ref{ccc}) is
\begin{eqnarray}\sum_{i=1,2}\int {\rm d}z \tilde\mu_i^{(1)} w_i'(z)
={\bar K}\sum_{i=1,2}\int {\rm d}z {\rm d}z' w'_i(z)(z-z')\tilde U(z-z')w_{j}(z').\label{s2}
\end{eqnarray}
which implies because $\tilde\mu_i^{(1)}$ are constant
\begin{eqnarray}\tilde\mu_1^{(1)}(0,r,t)[w_1]^{+\infty}_{-\infty}+\tilde\mu_2^{(1)}(0,r,t)[w_2]^{+\infty}_{-\infty}
={\bar K(r,t)}\sum_{i, i\neq j}\int {\rm d}z {\rm d}z' w'_i(z)(z-z')\tilde U(z-z')w_{j}(z'). \nonumber
\end{eqnarray}
In Appendix B it is shown that the  sum in the right hand side is the surface tension $S$ for this
model, so we have (since $[w_1]^{+\infty}_{-\infty}=-[w_2]^{+\infty}_{-\infty}$)
\begin{equation}(\tilde\mu_1^{(1)}-\tilde\mu_2^{(1)})(0,r,t)[w_1]^{+\infty}_{-\infty}= \bar K(r,t)S.
\label{s3}\end{equation} The matching conditions impose that $\tilde\mu_1^{(1)}-\tilde\mu_2^{(1)}\to
(\hat\mu_1^{(1)})^\pm-(\hat\mu_2^{(1)})^\pm$
 for $z\to\pm\infty$,  so
that  for $r\in\bar\Gamma_t$
\begin{equation}[(\hat\mu_1^{(1)})^\pm-(\hat\mu_2^{(1)})^\pm][w_1]^{+\infty}_{-\infty}=
\bar K(r,t)S. \label{s9}\end{equation}
and hence the continuity of $\mu_1^{(1)}-\hat\mu_2^{(1)}$ on the interface.

 The conservation law  for the equation at the order
$\e^{-1}$  ($n=2$) will give the velocity of the interface. By integrating over the velocity this equation we get
\begin{equation}
w' _i \bar V+  \partial_z (\bar\nu\cdot\tilde u_i^{(2)})=0 \ ,\label{s1}
\end{equation}
where the fact that  $\tilde f_i^{(0)}$ and $\tilde f_i^{(1)}$ are Maxwellian in velocity is crucial for several cancellations.
By integrating over $z$
\begin{equation}
 -\bar V [w_i]_{-\infty}^{+\infty}=[\bar\nu\cdot\tilde u_i^{(2)}]_{-\infty}^{+\infty}.
\end{equation}
By the matching conditions $\tilde u_i^{(2)}\to (\hat u_i^{(2)})^\pm$ at $\pm\infty$, so that for
$r\in\bar\Gamma_t$
\begin{equation}
 -\bar V [w_i]_{-\infty}^{+\infty}=[\bar\nu\cdot\hat u_i^{(2)}]^+_-\qquad
r\in\bar\Gamma_t. \label{velbulk}
\end{equation}
Summarizing what we got so far:  we have constructed functions $\hat \mu_i^{(1)}$  harmonic  in $\Omega^\pm$ which satisfy (\ref{s9}) and
(\ref{velbulk}). For sake of convenience we will denote by $\bar \mu_i^{(1)}$ the  functions defined in $\Omega$,  not necessarily  smooth, equal to
$(\hat
\mu_i^{(1)})^\pm$ in
$\Omega\setminus\bar\Gamma_t$ and such that $\lim_{d^{(0)}(r,t)\to 0^\pm}\bar \mu_i^{(1)}=(\hat
\mu_i^{(1)})^\pm|_{\bar\Gamma_t}$ and the same for the derivatives.   This means that $\bar \mu_i^{(1)}$ satisfy: $$\Delta \bar
\mu_i^{(1)}=0,\quad r\in\Omega\setminus\bar\Gamma_t,$$ $$(\bar \mu_1^{(1)}-\bar \mu_2^{(1)}) [\bar\rho^+-\bar\rho^-]=\bar
K(r,t)S,\quad r\in\Omega\setminus\bar\Gamma_t,$$
\begin{equation}
 \bar V [w_i]_{-\infty}^{+\infty}=[T\bar\rho_i\bar\nu\cdot\nabla_r \bar\mu_i^{(1)}]^+_- \label{vel}
\qquad r\in\bar\Gamma_t
\end{equation}
where  $\bar\rho^\pm=w_1(\pm \infty)$. Let us write the last equation as
\begin{equation}
\bar V \beta[\bar\rho^+-\bar\rho^-]=[(\bar\rho+\bar\varphi)\bar\nu\cdot\nabla_r \bar\mu_1^{(1)}]^+_-
=[(\bar\varphi-\bar\rho)\bar\nu\cdot\nabla_r \bar\mu_2^{(1)}]^+_-
\end{equation}
 and
$$\bar\rho(r)=\frac{\bar\rho_1(r)+\bar\rho_2(r)}{2},\quad \bar\varphi(r)=\frac{\bar\rho_1(r)-\bar\rho_2(r)}{2}.$$
with $ \bar\rho_i(r)$ the step functions $ \bar\rho_i(r):=\bar\rho^+_i\chi^++\bar\rho^-_i\chi^- $, $\chi^\pm$ the
characteristic functions of the sets ${d^{(0)}(r,t)>0}$, ${d^{(0)}(r,t)<0}$ respectively.   We know, because of the
symmetry of the phase transition, that $\bar\rho$ is constant while $\bar\varphi$ is discontinuous in $0$ and
$\bar\varphi(r)=\pm|\bar\phi|$ for $r\in\Omega^\pm$. The previous equation implies
\begin{equation}
2 \bar V \beta[\bar\rho^+-\bar\rho^-]=\bar\rho[\bar\nu\cdot\nabla_r (\bar \mu_1^{(1)}-\bar\mu_2^{(1)})]^+_-
+[\bar\varphi\bar\nu\cdot\nabla_r (\bar\mu_1^{(1)}+\bar\mu_2^{(1)})]^+_- \label{vel1}
\end{equation}
\begin{equation}\label{vel0}
 0=\bar\rho[\bar\nu\cdot\nabla_r (\bar\mu_1^{(1)}+\bar\mu_2^{(1)})]^+_-
+[\bar\varphi \bar\nu\cdot\nabla_r (\bar\mu_1^{(1)}-\bar\mu_2^{(1)})]^+_- .
\end{equation}
 We introduce the function $\zeta(r,t)
=(\bar\rho_1\bar\mu_1^{(1)}+\bar\rho_2\bar\mu_2^{(1)})(r,t)=\bar\rho (\bar\mu_1^{(1)}+\bar\mu_2^{(1)}) +\bar\varphi
(\bar\mu_1^{(1)}-\bar\mu_2^{(1)})$
 so that
$ \Delta_r\zeta(r, t)=0$ in $\Omega\setminus\bar\Gamma_t$ and (\ref{vel0}) gives $[\bar\nu\cdot\nabla_r
\zeta]^+_-=0$. Moreover, it is discontinuous on $\bar\Gamma_t$ because of the function $\bar\varphi$. The jump is
$$\zeta^+(r,t)-\zeta^-(r,t)=2|\bar\varphi|(\bar\mu_1^{(1)}-\bar\mu_2^{(1)}),\quad r\in \bar\Gamma_t$$ In
conclusion, $\zeta $ satisfies
\begin{equation}
\left\{
    \begin{array}{ll}
     \Delta_r\zeta(r, t)=0\quad &r\in \Omega\setminus\bar\Gamma_t \\
\displaystyle{[\zeta]^+_- =2|\bar\varphi| S \bar K(r,t)/[w_1]^{+\infty}_{-\infty}  }\qquad\qquad
  &r\in\bar\Gamma_t  \cr
  0 =\left[\bar\nu\cdot\nabla_r\zeta\right]_-^+ \qquad\qquad
&r\in\bar\Gamma_t
    \end{array}
\right.
\end{equation}
It is possible to show by using the Green identity that this problem   for a given function $\bar K(r,t)$, has the
unique solution $$ \zeta(r, t)= \int_{\bar\Gamma_t}{\rm d}s(\zeta^+-\zeta^-)(s,t)\nu\cdot\nabla
G(r,s)=\frac{2S|\bar\varphi|}{[w_1]^{+\infty}_{-\infty}}\int_{\bar\Gamma_t}{\rm d}s\bar K(s,t)\nu\cdot\nabla
G(r,s),\quad r\in\Omega\setminus\bar\Gamma_t$$
$$\frac{(\zeta^++\zeta^-)}{2}(r,t)=\frac{2S|\bar\varphi|}{[w_1]^{+\infty}_{-\infty}}\int_{\bar\Gamma_t}{\rm d}s\bar
K(s,t)\bar\nu\cdot\nabla G(r,s),\quad r\in\bar\Gamma_t$$where $G$ is the Green function in $\Omega$.  We notice
that $(\zeta^++\zeta^-)=2\bar\rho (\bar\mu_1^{(1)}+\bar\mu_2^{(1)})$.

We consider now the function $\xi(r,t)=(\bar\rho_1\bar \mu_1^{(1)}-\bar\rho_2\bar\mu_2^{(1)})(r,t)=\bar\rho
(\bar\mu_1^{(1)}-\bar\mu_2^{(1)}) +\bar\varphi (\bar\mu_1^{(1)}+\bar\mu_2^{(1)})$, which is discontinuous on
$\bar\Gamma_t$ and satisfies
\begin{equation}
\left\{
    \begin{array}{ll}
     \Delta_r\xi(r, t)=0 \qquad &r\in \Omega\setminus\bar\Gamma_t\\     \displaystyle{[\xi]^+_-
=\frac{|\bar\varphi|}{\bar\rho} (\zeta^++\zeta^-)}&r\in\bar\Gamma_t\\
  \bar V
=\frac{T}{2}\frac{\left[\bar\nu\cdot\nabla_r\xi\right]_-^+}{[\bar\rho^+-\bar\rho^-]}&r\in\bar\Gamma_t
    \end{array}
\right.
\end{equation}
 The problem is well posed because given the current configuration of the front the problem has a unique
solution and this solution in turn determines the velocity of the front.

 In conclusion we have determined $\bar\mu_1^{(1)}$ and $\bar\mu_2^{(1)}$. In
$\mathcal N^0(m)$ $\tilde \mu_i^{(1)}$ is constant  equal to the value  $\bar \mu_i^{(1)}(r,t),\quad
r\in\bar\Gamma_t$,  which is determined by solving the limiting equation. Hence  $\tilde\mu_i^{(1)}$ and $\hat
\mu_i^{(1)}$ are known at this stage. As a consequence, $\hat\rho_i^{(1)}$ are known through the relation $\hat
\mu_i^{(1)}=T\frac{\hat \rho_i^{(1)}}{\bar\rho_i}+\hat U\hat \rho_j^{(1)}$ in $\Omega\setminus \mathcal N^0(m)$ while
$\tilde\rho_i^{(1)}$ are found as solutions of (\ref{ccc})
 with the r.h.s.   decaying to a constant as $z\to\pm\infty$ and the decay is exponential if
$w_i$ do so. Then, a modification of the argument in \cite{CCO1} leads to the exponential decay of  $\tilde
\rho_i^{(1)}$. We notice that $\tilde\rho_i^{(1)}$ is determined by (\ref{ccc}) but for a term $\a w'_i$  which is
in the null of $\mathcal L$, with $\a$ independent of $z$. To fix $\a$ it is enough to put the condition
$\tilde\rho_1^{(1)}(0,r,t))=\tilde\rho_2^{(1)}(0,r,t),\quad r\in \mathcal N^0(m)$.  Since we have fixed
$\rho^\e_1=\rho^\e_2$ on $\Gamma^\e$ we are allowed to choose
$\tilde\rho_1^{(k)}(0,r,t))=\tilde\rho_2^{(k)}(0,r,t),\quad r\in \mathcal N^0(m)$ for any $k$.

We proceed now constructing the higher orders of the expansion.
 For  $n=2$:
\begin{eqnarray}
 &\displaystyle{\bar V\partial_z  \tilde f_i^{(0)}+\bar\nu\cdot v\partial_z \tilde f_i^{(2)}+
 v\cdot\overline{\nabla}_r  \tilde f_i^{(1)}-\partial_z \tilde g_i^{(2)}\bar\nu\cdot\nabla_v
\tilde f_i^{(0)}-\partial_z \tilde g_i^{(0)}\bar\nu\cdot\nabla_v \tilde f_i^{(2)} }&\nonumber \\
 &\displaystyle{-\partial_z \tilde g_i^{(1)}\bar\nu\cdot\nabla_v
\tilde f_i^{(1)} +\overline{\nabla}_r \tilde g_i^{(1)}\cdot\nabla_v \tilde f_i^{(0)}=L_{\beta} \tilde
f_i^{(2)}}.&\label{bbb}
\end{eqnarray}
Again, the terms involving $\nu^{(2)}$ and  $\nu^{(1)}$ are zero thanks to the previous equations. The matching
conditions require for $z$ large $$\tilde f_i^{(2)}(\pm|z|,r,t) =(\hat f_i^{(2)})^\pm+\bar\nu\cdot(\nabla_r \hat
f_i^{(1)})^\pm(z-d^{(1)}) +O(e^{-\a |z|}).$$
 Hence, we have to solve  a  stationary problem on the real line  with given
conditions at infinity.
 We replace in
(\ref{bbb}) $ \tilde f_i^{(2)} = \tilde q_i^{(2)}+ \tilde\rho^{(2)}_i M_{\beta}$ with $\int dv  \tilde q_i^{(2)}=0$. This
means that $\tilde q_i^{(2)}$ is in the orthogonal to the kernel of $L_{\beta}$ versus the scalar product 
\[
(f,g)_{M_{\beta}}=\int dvM_{\beta}^{-1}fg
\]
We have
\begin{eqnarray}
 &\displaystyle{M_\beta[\bar V\partial_z  w_i+\bar\nu\cdot v\partial_z  \tilde \rho_i^{(2)}
+ v\cdot\overline{\nabla}_r \tilde \rho_i^{(1)}+\beta \partial_z \tilde g_i^{(2)}\bar\nu\cdot v w_i+\beta\partial_z
\tilde g_i^{(0)}\bar\nu\cdot v \tilde \rho_i^{(2)}+\beta
\partial_z \tilde g_i^{(1)}\bar\nu\cdot v
\tilde \rho_i^{(1)} }&\nonumber \\  &\displaystyle{ -\beta v\cdot \overline{\nabla}_r \tilde g_i^{(1)}
w_i]=L_{\beta} \tilde q_i^{(2)}-\bar\nu\cdot v\partial_z   \tilde q_i^{(2)}+\partial_z \tilde
g_i^{(0)}\bar\nu\cdot\nabla_v \tilde q_i^{(2)}}.& \label{inner2}
\end{eqnarray}
By using the equation for the front $w_i$ and the fact that $\partial_z \tilde \mu_i^{(1)}=0$ together with
$\tilde\mu_i^{(2)}=\frac{T}{w_i}\tilde\rho_i^{(2)} -\frac {T}{2}(\frac{\tilde\rho_i^{(1)}}{w_i})^2+\tilde
g_i^{(2)}$ we get $$M_\beta\bar\nu\cdot v[\partial_z \tilde\rho_i^{(2)}+\beta w_i\partial_z \tilde
g_i^{(2)}+\beta\tilde \rho_i^{(2)}\partial_z \tilde g_i^{(0)}+\beta
\partial_z \tilde g_i^{(1)}
\tilde \rho_i^{(1)}]$$ $$=M_\beta\bar\nu\cdot v w_i\partial_z[\frac{\tilde\rho_i^{(2)}}{w_i}+\beta \tilde
g_i^{(2)}- \frac {\beta}{2}(\frac{\tilde\rho_i^{(1)}}{w_i})^2] =\beta M_\beta\bar\nu\cdot v
w_i\partial_z\tilde\mu_i^{(2)}$$
 Hence we can write the equation (\ref{inner2}) in the form
 \goodbreak
\begin{eqnarray}\beta M_\beta\bar\nu\cdot v w_i\partial_z\tilde\mu_i^{(2)}&=&L_{\beta} \tilde
q_i^{(2)}-\bar\nu\cdot v\partial_z   \tilde q_i^{(2)}+\partial_z \tilde g_i^{(0)}\bar\nu\cdot\nabla_v \tilde
q_i^{(2)}+\beta M_{\beta}w_iv\cdot\overline{\nabla}_r\tilde g_i^{(1)}-\cr &-&M_\beta v\cdot\overline{\nabla}_r
\tilde \rho_i^{(1)}-M_\beta\bar V\partial_z w_i \label{f3B}
\end{eqnarray}
From
 (\ref{f22})  the conditions at infinity are:
$$\tilde
f_i^{(2)}(\pm|z|r,t)=M_{\beta}\left[(\hat\rho_i^{(2)})^{\pm}-(\hat\rho_i^{(0)})^\pm v\cdot(\nabla_r\hat\mu_i^{(1)})^{\pm}+
\bar\nu\cdot(\nabla_r\hat\rho_i^{(1)})^\pm(z-d^{(1)})\right] +O(e^{-\a |z|})$$ $$\tilde
\rho_i^{(2)}(\pm|z|r,t) = (\hat
\rho^{(2)}_i)^\pm(r,t)+\bar\nu\cdot(\nabla_r\hat\rho_i^{(1)})^\pm(z-d^{(1)})+O(e^{-\a
|z|})$$ $$\lim_{z\to\pm\infty}\int dv P(v)\tilde q_i^{(2)}(z,r,t)=-<vP(v)>_\beta
\cdot(\nabla_r\hat\mu^{(1)}_i)^\pm(\hat\rho^{(0)}_i)^\pm$$ where $P(v)$ is a polynomial in the velocity and
$<\cdot>_\beta$ are the moments of the Maxwellian $M_\beta$: $<vP(v)>_{\beta}=\int dvvP(v)M_{\beta}$. The matching conditions for the chemical
potential imply that as $z\to\pm\infty$
\begin{eqnarray*}
 &\displaystyle{\tilde\mu_i^{(2)}=(\hat\mu_i^{(2)})^\pm+
 (z-d^{(1)})\bar\nu\cdot(\nabla_r\mu_i^{(1)})^\pm+O(e^{-\a |z|})}:=(\hat\mu_i^{(2)})^\pm+C_i^\pm&
\end{eqnarray*}
We have that
\[
w_i(+\infty)(\hat\mu_i^{(2)})^+-w_i(-\infty)(\hat\mu_i^{(2)})^-=[w_i(\tilde\mu_i^{(2)}-C_i)]_{-\infty}^{+\infty}
\]
where $
C_i=1_{z<0}C^-_i+1_{z>0}C^+_i$. 
The left hand side can be written as
\[
\int dz\partial_z[w_i(\tilde\mu_i^{(2)}-C_i)]=\int
dz[w_i\partial_z(\tilde\mu_i^{(2)}-C_i)+w_i'(\tilde\mu_i^{(2)}-C_i)]
\]
We multiply (\ref{f3B}) by $v_z=\bar\nu\cdot v$
and integrate over $v$
\begin{equation}\label{velnorm}
w_i\partial_z\tilde\mu_i^{(2)}=-\partial_z\int dv v_z^2 \tilde q_i^{(2)} -\beta\bar\nu\cdot\tilde u_i^{(2)}
\end{equation}
where we used $\tilde q_i^{(2)}\in[\mbox{Ker}L_{\beta}]^{\perp}$, $\bar\nu\cdot\bar\nabla(\cdot)=0$ and 
$L_{\beta}\tilde q_i^{(2)}=L_{\beta}\tilde f_i^{(2)}$. We have also 
\begin{equation}
w_i\partial_z(\tilde\mu_i^{(2)}-C_i)=-\partial_z\int dvv_z^2\tilde q_i^{(2)}-\beta\bar\nu\cdot\tilde
u_i^{(2)}-w_i\partial_zC_i
\label{CCC}
\end{equation}
By matching conditions we have that for $z$ large $$
\beta\bar\nu\cdot\tilde u_i^{(2)}(\pm|z|r,t)=-(\hat\rho_i^{(0)})^\pm \bar\nu\cdot(\nabla_r\hat\mu_i^{(1)})^{\pm}+
O(e^{-\a |z|})$$ 
and that $\int dv v_z^2\tilde q_i^{(2)}$  vanishes at infinity.
Since 
\[
\partial_zC_i=1_{z<0}(\bar\nu\cdot\nabla_r\hat\mu_i^{(1)})^-+1_{z>0}(\bar\nu\cdot\nabla_r\hat\mu_i^{(1)})^+
\]
we have that asymptotically in $|z|$
$$\beta\bar\nu\cdot\tilde
u_i^{(2)}+w_i\partial_zC_i=O(e^{-\a|z|})$$
so that  the integral over $z$ of  the r.h.s. of (\ref{CCC}) makes sense.
Then
\begin{eqnarray}
 &
w_i(+\infty)(\hat\mu_i^{(2)})^+-w_i(-\infty)(\hat\mu_i^{(2)})^-\qquad\qquad\qquad\qquad\cr
 &=-\int dz[\beta\bar\nu\cdot\tilde
u_i^{(2)}+w_i\partial_zC_i]+\int dzw_i'\tilde\mu_i^{(2)}-\int dzw_i'C_i
\label{98}
\end{eqnarray}
 Moreover the function
$\bar\nu\cdot\tilde u_i^{(2)}$ is known because
\[
\bar Vw_i'+\partial_z\bar\nu\cdot\tilde u_i^{(2)}=0
\]
It follows that
\[
\bar\nu\cdot\tilde u_i^{(2)}=\frac{w_i-\bar\rho^-}{\bar\rho^+-\bar\rho^-}\left[(\frac{\hat\rho_i^{(0)}}{\beta}
\bar\nu\cdot\nabla_r\hat\mu_i^{(1)})^--(\frac{\hat\rho_i^{(0)}}{\beta}
\bar\nu\cdot\nabla_r\hat\mu_i^{(1)})^+\right]-(\frac{\hat\rho_i^{(0)}}{\beta}
\bar\nu\cdot\nabla_r\hat\mu_i^{(1)})^-
\]
So only the last term involving $\tilde\mu_i^{(2)}$ in (\ref{98}) needs to be computed. But we know that
\begin{equation}
\tilde\mu_i^{(2)}=\frac{T}{w_i}\tilde\rho_i^{(2)} -\frac {T}{2}(\frac{\tilde\rho_i^{(1)}}{w_i})^2+\tilde
g_i^{(2)}=\mathcal L \tilde \rho_i^{(2)} +B_1(\tilde\rho_i^{(1)})+B_2(\tilde\rho_i^{(0)}) . \label{inv}
\end{equation}
where $B_1$ and $B_2$ are defined in Appendix C. 
The solvability condition is
\begin{equation}\sum_i\int_{-\infty}^{+\infty} dz[\tilde\mu_i^{(2)}- D_i]w_i'=0 ,
\label{con}
\end{equation}
where $D_i:=B_1(\tilde\rho_i^{(1)})+B_2(\tilde\rho_i^{(0)})$. 
Summing on $i=1,2$
\begin{eqnarray}
&\sum_i\int dzw_i'\tilde\mu_i^{(2)}=\sum_i\int_{-\infty}^{+\infty} dz D_iw_i'=\sum_i\int_{-\infty}^{+\infty}
dz w_i'(B_1(\tilde\rho_i^{(1)})+B_2(\tilde\rho_i^{(0)}))
\end{eqnarray}
We introduce functions $\bar \mu_i^{(2)}$ defined  as explained after (\ref{velbulk}). We have
\begin{equation}
[\bar\rho_1\bar\mu_1^{(2)}+\bar\rho_2\bar\mu_2^{(2)}]^+_-=\sum_i\big[-\int dz[\beta\bar\nu\cdot\tilde
u_i^{(2)}+w_i\partial_z C_i]+\int dz w'_i(D_i-C_i)\big]=:H
\label{???}
\end{equation}
We are now in position to find the first correction to the velocity of the interface, $V^{(1)}$. This is given by
the solvability condition for the boundary equation
 for
$n=3$:
\begin{eqnarray*}
& \displaystyle{\sum_{l,l'\geq 0: l+l'=1}V^{(l')}\partial_z  \tilde f_i^{(l)}+\sum_{l,l'\ge
0:l+l'=3}v\cdot\nu^{(l)}\partial_z\tilde f_i^{(l')}+ v\cdot\overline{\nabla}_r \tilde f_i^{(2)}+\partial_t \tilde
f_i^{(0)}- }&\\
 &\displaystyle{ -\sum_{l,l',l''\geq 0: l+l'=3}\partial_z \tilde g_i^{(l)}\nu^{(l'')}\cdot\nabla_v \tilde f_i^{(l')} -\sum_{l,l'\geq 0:
l+l'=2}\overline{\nabla}_r \tilde g_i^{(l)}\cdot\nabla_v \tilde f_i^{(l')}=L_{\beta} \tilde f_i^{(3)}}&
\end{eqnarray*}
After integration on $v$ we get $$\overline\nabla_r\cdot\tilde u_i^{(2)}+\bar V\partial_z \tilde\rho^{(1)}_i
+V^{(1)}\partial_z w_i +
\partial_z (\bar\nu\cdot \tilde u_i^{(3)})+\partial_z (\nu^{(1)}\cdot \tilde u_i^{(2)})=0 $$
Taking into account that for $z$ large 
\goodbreak
$$\overline{\nabla}_r\cdot\tilde u_i^{(2)}+\partial_z \bar\nu\cdot \tilde u_i^{(3)}=
(\overline{\nabla}_r\cdot\hat u_i^{(2)})^\pm+ \bar\nu\cdot (\nabla_r(\bar\nu\cdot\hat u_i^{(2)}))^\pm+O(e^{-\a
|z|})$$ $$\qquad\qquad\quad=(\nabla_r\cdot\hat u_i^{(2)})^\pm+O(e^{-\a |z|})=O(e^{-\a |z|}),$$ 
because $\nabla_r
\cdot\hat u_i^{(2)}=0$, we can integrate over $z$:
\begin{equation}
\bar V[\tilde\rho^{(1)}_i]^{+\infty}_{-\infty}+V^{(1)}[w_i]^{+\infty}_{-\infty}=-\int
dz[\overline{\nabla}_r\cdot\tilde u_i^{(2)}+\partial_z(\bar\nu\cdot\tilde u_i^{(3)})]-[\nu^{(1)}\cdot \tilde
u_i^{(2)}]^{+\infty}_{-\infty}.
\end{equation}
By the matching conditions we have as $z\to\pm\infty$
\[
\tilde u_i^{(3)}\sim(\hat u_i^{(3)})^\pm+(z-d^{(1)})\bar\nu\cdot(\nabla_r\hat u_i^{(2)})^\pm
\]
Let us introduce the functions
\[
n_i=1_{z<0}(z-d^{(1)})\bar\nu\cdot(\nabla_r(\bar\nu\cdot\hat u_i^{(2)}))^-+1_{z>0}
(z-d^{(1)})\bar\nu\cdot(\nabla_r(\bar\nu\cdot\hat u_i^{(2)}))^+
\]
so that 
\[
\partial_zn_i=1_{z<0}\bar\nu\cdot(\nabla_r(\bar\nu\cdot\hat
u_i^{(2)}))^-+1_{z>0}\bar\nu\cdot(\nabla_r(\bar\nu\cdot\hat u_i^{(2)}))^+
\]
Thus
\begin{eqnarray*}
 &\displaystyle{\int dz [\overline{\nabla}_r\cdot\tilde u_i^{(2)}+\partial_z(\bar\nu\cdot\tilde u_i^{(3)})]= \int
 dz[\overline{\nabla}_r\cdot\tilde u_i^{(2)}+\partial_zn_i+\partial_z(\bar\nu\cdot\tilde u_i^{(3)}-n_i)]=}&\\
 &\displaystyle{=\int dz[\overline{\nabla}_r\cdot\tilde u_i^{(2)}+\partial_zn_i]+[\bar\nu\cdot\hat u_i^{(3)}]_-^+=A_i+[\bar\nu\cdot\hat
u_i^{(3)}]_-^+}&
\end{eqnarray*}
where we still have to compute $A_i$. Essentially we need to know $\overline{\nabla}_r\cdot\tilde u_i^{(2)}$. It
can be derived from (\ref{f3B}) multiplying for $\bar\nu_{\perp}\cdot v$, where $\bar\nu_{\perp}$ denotes one of the two
directions orthogonal to $\bar\nu$, and integrating in $v$. After some computations we have
\[
\beta\bar\nu_{\perp}\cdot\tilde u_i^{(2)}=-\partial_z\int dv(\bar\nu_{\perp}\cdot v)(\bar\nu\cdot v)\tilde
q_i^{(2)}+w_i\bar\nu_{\perp}\cdot\overline{\nabla}_r\tilde
g_i^{(1)}+\frac{1}{\beta}\bar\nu_{\perp}\cdot\overline{\nabla}_r\tilde\rho_i^{(1)}
\]
In order to obtain the overlined divergence of $\tilde u_i^{(2)}$ we have to sum over the two orthogonal
directions; then remembering the asymptotic behaviour of $\int dv P(v)\tilde q_i^{(2)}$, we can conclude that
\[
A_i=\int dz\left[\frac{1}{\beta}w_i\overline{\nabla}_r\cdot\overline{\nabla}_r\tilde g_i^{(1)}+\frac{1}{\beta^2}
\overline{\nabla}_r\cdot\overline{\nabla}_r\tilde\rho_i^{(1)}+\partial_zn_i\right]
\]
Here we recall that $\hat u_i^{(2)}$ depends only on quantities of the previous order in $\e$, which are known.
Moreover by (\ref{f3}) we have also $$\hat u_i^{(3)}=-  T[\bar\rho_i\bar\nu\cdot\nabla_r\hat\mu^{(2)}_i+
\bar\rho^{(1)}_i\bar\nu\cdot\nabla_r \hat\mu^{(1)}_i]$$ so that
\begin{equation}\label{V1}
\frac{1}{\beta}[\hat\rho_i^{(0)}\bar\nu\cdot\nabla_r\hat\mu_i^{(2)}]_-^+=\bar
V[\hat\rho_i^{(1)}]_-^++V^{(1)}[\hat\rho_i^{(0)}]_-^+-\frac{1}{\beta}[\hat\rho_i^{(1)}\bar\nu\cdot\nabla_r\hat
\mu_i^{(1)}]_-^++A_i-\frac{1}{\beta}[\hat\rho_i^{(0)}\nu^{(1)}\cdot\nabla_r\hat\mu_i^{(1)}]_-^+
\end{equation}
 We consider the functions $\zeta^{(2)}=\bar\rho_1\bar \mu_1^{(2)}+\bar\rho_2\bar
\mu_2^{(2)}$ and $\xi^{(2)}=\bar\rho_1\bar \mu_1^{(2)}-\bar\rho_2\bar\mu_2^{(2)}$. We have from (\ref{???})
$$[\zeta^{(2)}]^+_-=H(r,t)$$ and by summing in equation (\ref{V1})
\[
[\bar\nu\cdot\nabla_r\zeta^{(2)}]_-^+=\beta\bar
V[\hat\rho_1^{(1)}+\hat\rho_2^{(1)}]_-^++\sum_i\left\{-[\hat\rho_i^{(1)}\bar\nu\cdot\nabla_r\hat\mu_i^{(1)}]_-^++\beta
A_i-[\hat\rho_i^{(0)}\nu^{(1)}\cdot\nabla_r\hat\mu_i^{(1)}]_-^+\right\}:=P
\]
Moreover, we  have the identity $[\xi^{(2)}]^+_-=\zeta^{(2)+}+\zeta^{(2)-}$.

We get the velocity $V^{(1)}$  by taking the difference in (\ref{V1}) on the index $i$
\begin{eqnarray*}
 &\displaystyle{2[\bar\rho^+-\bar\rho^-]V^{(1)}=\frac{1}{\beta}[\bar\nu\cdot\nabla_r\xi^{(2)}]_-^+-\bar
 V[\hat\rho_1^{(1)}-\hat\rho_2^{(1)}]_-^++\frac{1}{\beta}[\hat\rho_1^{(1)}\bar\nu\cdot\nabla_r\hat\mu_1^{(1)}]_-^+-}&\\
 &\displaystyle{-\frac{1}{\beta}[\hat\rho_2^{(1)}\bar\nu\cdot\nabla_r\hat\mu_2^{(1)}]_-^+-A_1+A_2+\frac{1}{\beta}
 [\hat\rho_1^{(0)}\nu^{(1)}\cdot\nabla_r\hat\mu_1^{(1)}]_-^+-}&\\
 &\displaystyle{-\frac{1}{\beta}[\hat\rho_2^{(0)}\nu^{(1)}\cdot\nabla_r\hat\mu_2^{(1)}]_-^+:=2[\bar\rho^+-\bar\rho^-]Q}&
\end{eqnarray*}
Notice that $\int _{\Gamma_t} V^{(1)}$ is not necessarily zero as was $\int _{\Gamma_t} \bar V$. This implies that
the volume enclosed by   the interface ${\Gamma_t^{(1)}} $ evolving with $V^{(0)}+\e V^{(1)}$ is not conserved.

 In conclusion,  $\xi^{(2)}$ and $\zeta^{(2)}$ are solutions of
\begin{equation}
\left\{
    \begin{array}{ll}
     \Delta_r\xi^{(2)}(r, t) =  (S_1-S_2)(r, t)\qquad & r\in \Omega\setminus\bar \Gamma_t\cr
     [\xi^{(2)}]^+_- =
\frac{|\bar\phi|}{ \bar\rho}(\zeta^{{(2)}+}+\zeta^{{(2)}-})\qquad & r\in \bar \Gamma_t \\ V^{(1)}=Q & r\in \bar
\Gamma_t\\
\end{array}
\right . \label{MSe}
\end{equation}
  and
\begin{equation}
\left\{
\begin{array}{ll}
\Delta_r\zeta^{(2)}(r,t)=(S_1+S_2)(r, t)\quad & r\in\Omega\setminus\bar\Gamma_t\\ \displaystyle{[\zeta^{(2)}]^+_-=H(r,t)
}&r\in\bar\Gamma_t\\ \left[\bar\nu\cdot\nabla_r\zeta^{(2)}\right]_-^+=P(r, t)& r\in\bar\Gamma_t
\end{array}
\right. \label{HSh}
\end{equation}
$S_i, Q$ have been determined before. The terms $H$ and $P$ depend  on $d^{(0)}$ which has been already found  and
also on $d^{(1)}$ which is unknown and has to be determined by $V^{(1)}=\partial_t d^{(1)}$. These equations are
different from the first order equations because the surface $\bar \Gamma_t$ is given, so that we are not facing a
free boundary problem. In this sense they are ``linearized" even if the equations remain non linear. The problem
is well posed  because given $d^{(1)}$ on $\bar \Gamma_t$ the problem has a unique solution and this solution in
turn determines the velocity $V^{(1)}$. Then, $d^{(1)}$ is found in $\mathcal N^0(m)$ through the condition $\nabla
d^{(1)}\nabla d^{(0)}=0$. The argument is analogous to the one in \cite{ABC}.

Once $\bar \mu_i^{(2)}$ are found as solutions of these equations  we can find $\hat\rho_i^{(2)}$ in terms of $\hat
\mu_i^{(2)}$.  We have now the asymptotic values needed to solve (\ref{f3B}).  If the solution  exists, it decays
exponentially at infinity, because the known terms have this property. This equation admits a solution if the
conditions at infinity satisfy suitable conditions. The matching conditions require that the solution at infinity
grows linearly. This is a problem analogous to the so-called Kramers problem in the half space \cite{BCN}.  It can
be reduced to a Riemann problem with fixed asymptotic values at infinity in the following way:

\noindent Since the solution $\tilde f_i^{(2)}$ has to be approximately $A_i^\pm+zB^\pm_i$ at infinity, the
functions $A_i^\pm=M_{\beta}((\hat \rho^{(2)}_i)^\pm- v\cdot(\hat\rho^{(0)}_i\nabla_r\hat\mu^{(1)}_i )^\pm)$ and
$B_i^\pm=M_\beta(\nu\cdot \nabla_r \hat\rho_i^{(1)})^\pm$ have to satisfy $$\nu\cdot v (B_i^\pm + M_\beta\beta \hat
U (\hat\rho_i^ {(0)}\bar\nu\cdot\nabla_r \hat\rho_i^{(1)})^\pm)+M_\beta
v\cdot(\hat\rho^{(0)}_i\overline\nabla_r\hat\mu_i^{(1)})^\pm=L_\beta(A_i^\pm+zB_i^\pm) $$ because all the other
terms in (\ref{bbb}) vanish at infinity. This is equivalent to $$L_\beta(B_i^\pm)=0,\qquad L_\beta A_i^\pm=
\bar\nu\cdot v ( B_i^\pm + M_\beta\beta \hat U (\hat\rho_i^{(0)}\nu\cdot\nabla_r\hat\rho_i^{(1)})^\pm) +M_\beta
v\cdot(\hat\rho^{(0)}_i\overline\nabla_r\hat\mu_i^{(1)})^\pm.$$ This is true  by direct inspection. Then, the
problem is reduced to a well posed problem of finding a solution $h_i$ to eq. (\ref{bbb}) decaying to
$M_{\beta}(\hat \rho^{(2)}_i)^\pm$ at infinity.

  Similar arguments lead to the computation of higher order terms.

We conclude this section by remarking that our expansion is different from the one in \cite{CCO1} which is more
similar to a Chapmann-Enskog expansion because the terms of their expansion $f_i^{(n)}$ still depend on $\e$ and
are determined by equations which are nonlinear in the interface at every order in the sense that they are
free-boundary problems determining for any $n$ an interface $\Gamma^{(n)}$ moving with velocity $\sum_{i=0}^n
\e^iV^{(i)}$.  Our approach is still based on the matching conditions and is in a way intermediate between
\cite{ABC} and
\cite{Yu}, where it is proven the hydrodynamic limit for the Boltzmann equation in presence of shocks by
constructing a Hilbert expansion to approximate the solution.

\bigskip

\section{Interface motion}
\setcounter{equation}{0}

In this section we  discuss the equations for the interface motion. We start by rewriting them in terms of the
quantities $\zeta$ and $\psi=\bar\mu_1^{(1)}-\bar\mu_2^{(1)}$. The equation for $\psi$ is similar to the
Mullins-Sekerka equation   but for the fact that there is an extra term determining the velocity 
\begin{equation}
\left\{
    \begin{array}{ll}
     \Delta_r\psi(r, t) = 0 \qquad  & r\in \Omega\setminus\Gamma_t \\   \displaystyle{\psi(r,t)
= \frac{S \bar K(r,t)}{\bar\rho^+-\bar\rho^-}  }&r \in \Gamma_t\\
V =\frac{T}{2[\bar\rho^+-\bar\rho^-]}\Big[\frac{1}{{\bar\rho}}
(\bar\rho^2-|\bar\varphi|^2){\left[\bar\nu\cdot\nabla_r\psi\right]_-^+}
+ \frac{1}{\bar\rho}\left[{\bar\varphi}\bar\nu\cdot\nabla_r\zeta\right]_-^+\Big]&r \in \Gamma_t\\
\end{array}
\right .
\label{MS}
\end{equation}
 The  jump of ${\bar\varphi}\bar\nu\cdot\nabla_r\zeta$ in the last term on the r.h.s. is indeed
$2|\bar\varphi|
\nu\cdot\nabla_r\zeta(r,t), r\in
\Gamma_t$ and
\begin{equation}
\left\{
    \begin{array}{ll}
     \Delta_r\zeta(r, t)=0\qquad  &r\in \Omega\setminus\Gamma_t\\  \displaystyle{[\zeta]^+_-
=2|\bar\varphi| S \bar K(r,t)/[w_1]^{+\infty}_{-\infty} }\qquad &r \in \Gamma_t\\
  0 =\left[\bar\nu\cdot\nabla_r\zeta\right]_-^+&r \in \Gamma_t
    \end{array}
\right.
\label{HS'}
\end{equation}
 Hence there are two contributions to the velocity of the interface: $V_{MS}$, the velocity of an interface in
the Mullins-Sekerka motion,  and $V_{HS}$, the velocity of an interface in the two-phases Hele-Shaw motion
(\ref{HS'}). The latter describes the motion of a bubble of gas expanding into a fluid in a radial Hele-Shaw cell
and is a free-boundary problem for the pressure $P$
\begin{equation}
\left\{
    \begin{array}{ll}
     \Delta_r P(r, t)=0\quad &r\in \Omega\setminus\Gamma_t\\  \displaystyle{[p]^+_-
= C  K(r,t)/[w_1]^{+\infty}_{-\infty} }& r\in \Gamma_t\\
  V =\nu\cdot\nabla_rP& r\in \Gamma_t
    \end{array}
\right.
\end{equation}
 Equations \ref{MS} and \ref{HS'} are identical to the equations in \cite{OE}, describing the sharp interface
motion for the dynamics of incompressible fluid mixtures driven by thermodynamic forces, modeling a polymer blend.
In this paper the macroscopic equation is a modification of the Cahn-Hilliard equation for a mixture of two fluids
that includes a lagrangian multiplier $p$ ("pressure") to take into account  the constraint of constant total
density
\begin{eqnarray}
&&\partial_t\rho_i =\nabla\cdot(\rho_i \nabla(\mu_i+p))\quad i=1,2\cr
&&\rho_1+\rho_2=1 .
\end{eqnarray}

This produces in the macroscopic
equation for the concentration a convective term which in turn gives rise to the Hele-Shaw contribution
$V_{HS}$ to the interface motion. The macroscopic equations (\ref{0.4}) with
$\mu_i=\delta\mathcal{F}/\delta\rho_i$ differs from the ones above for the constraint and hence for the
pressure term.
 It is easy to
see that the formal sharp interface limit  is the same for both equations with  $\nabla
\zeta$
 in the bulk a divergence-free field appearing as a velocity  field in the equation for the total density which
is constant in the bulk at the first order. Moreover, thermodynamic relations give that $\nabla \zeta=\nabla
p^{(1)}$ with $p^{(1)}$ the first correction to the  effective   pressure. Hence, the role of $\nabla \zeta$
is exactly the same as the lagrangian multiplier $p$ in \cite{OE}.

We refer  to \cite{OE} for the discussion on the behavior of the interface as given in (\ref{MS}) and
(\ref{HS'}).  Here we want just to  remark  that the Hele-shaw motion has more conserved quantities than the
Mullins-Sekerka motion. In fact, the former  conserves the volume of each connected component of both phases,
while the latter conserves only the total volume as we can easily see by starting from
$$\frac{d}{dt} |\Omega^+_{\Gamma}|=\int _{\Gamma} V$$
where $\Omega^+_{\Gamma}$ is the region enclosed in the surface $\Gamma$.
We consider now  a situation in which there are $N$ closed curves $\Gamma_i$ dividing $\Omega$ in $N$ connected
components $\Omega^+_{\Gamma_I}$. In the Mullins-Sekerka problem the velocity is proportional to the jump of the
normal derivative of the harmonic function
$f$ and this implies by using the divergence theorem
$$\sum_i\frac{d}{dt} |\Omega^+_{\Gamma_i}|=\sum_i\int _{\Gamma_i} [\nu\cdot
\nabla f]^+_-=\sum_i\int_{\Omega^+_{\Gamma_i}}\Delta f+\int_{\Omega^-}\Delta f =0$$
where $\Omega^-$ is the complement of $\cup_i (\Omega^+_{\Gamma_i}\cup \Gamma_i)$.
In the Hele-Shaw problem  the velocity is proportional to the
normal derivative of $f$ and
$$\frac{d}{dt} |\Omega^+_{\Gamma_i}|=\int _{\Gamma_i} \nu\cdot\nabla
f=\int_{\Omega^+_{\Gamma_i}}\Delta f =0$$

In the problem \ref{MS}, \ref{HS'} this fact  has consequences on the evolution of the droplets of
the two phases. The relative importance of the two contributions
$V_{HS}$ and
$V_{MS}$ is ruled by the coefficients: if $(\bar\rho^-)^{-1}-(\bar\rho^+)^{-1}<<1$ (near the critical point) the
$V_{MS}$ term dominates, while for deep quenches   the $V_{HS}$ term prevails.

\bigskip

\appendix{}

\section{{\it Uniqueness}}
\setcounter{equation}{0}

In this Appendix we prove that the equations for the $\tilde f^{(n)}(z,r,t)$ we examined in section 5 have solutions
whose dependence on the velocity is necessarily gaussian. We will omit for simplicity the dependence on the other
variables.

We consider  the following set of equations for $h_i(z,v),\quad i=1,2,\quad (z,v)\in \R\times \R^d$
\begin{equation}
v_z\partial_z h_i +F_i\partial_{v_z}h_i=  L_\beta h_i
\label{order0}
\end{equation}
where  $F_i=-\partial_z\int dz' U(|z-z'|)\rho_j(z'):=-\partial_z V_i, \quad i\neq j$, $\rho_i(z)=\int dv h_i(z,v)$,
with the conditions at infinity
\begin{equation}
h_i(\pm\infty,v)=M(v)\rho_i^\pm
\label{BC}
\end{equation}
 and show that it
has only a solution of the form $M(v)\rho_i(z)$.

Put $ h_i=\psi_i(z,v)M(v)e^{-\beta V_i}$.  $V_i$ is bounded due to the assumptions on $U$.
Then,
\begin{equation}
v_z\partial_z\psi_i +F_i\partial_{v_z}\psi_i=\tilde L_i\psi_i
\end{equation}
where
\[
\tilde L\psi_i=\frac{1}{M_{\beta}}\nabla_v\cdot(M_{\beta}\nabla_v\psi_i)
\] with the conditions at infinity
$$ \psi_i(\pm\infty,v)=e^{+\beta V_i(\pm\infty)}\rho_i^\pm
$$
 Multiply by
$M_\beta\psi_i$ and integrate over
$v$
\begin{equation}
\frac{1}{2}\partial_z(v_z\psi_i,\psi_i)_{M_\beta} +F_i(\psi_i,\frac{d}{dv_z}\psi_i)_{M_\beta}=-(\psi_i,\tilde
L\psi_i)_{M_\beta}
\end{equation}
where $(h,g)_{M_\beta}=\int dvh(v)g(v)M_\beta(v)$. We have
\begin{equation}
\frac{1}{2}\frac{d}{dz}(v_z\psi_i,\psi_i)_{M_\beta}
-\frac{\beta}{2}F_i(v_z\psi_i,\psi_i)_{M_\beta}=-e^{-{\beta}
V_i}(\nabla_v\psi_i,\nabla_v\psi_i )_{M_\beta}
\end{equation}
that we write as
\begin{equation}
\frac{d}{dz}[(v_z\psi_i,\psi_i)_{M_\beta}e^{-{\beta} V_i}]
=-2(\nabla_v\psi_i,\nabla_v\psi_i )_{M_\beta}e^{\beta V_i}
\end{equation}
We notice that $(v_z\psi_i,\psi_i)(\pm\infty)=0$ because of the boundary conditions. Hence, by integrating over $z$
we get  $$\int_{-\infty}^{+\infty} {\rm d}ze^{-{\beta} V_i} ||\nabla_v\psi_i||_\beta^2=0 $$
which implies $\nabla_v\psi_i=0 $ a.e. since $V_i$ is bounded, so that $\psi_i=g(z)$, a function only of the
position. Going back to the original equation we see that $g(z)$ has to be the front solution. \vskip.2cm
 Next order equation.

We discuss now  equation (\ref{aab}).
  A solution has been
explicitly found as a Maxwellian times the density $\tilde\rho^{(1)}$. Suppose that there are two different solutions
$h$ and $h'$ such that
$\rho_ h=\rho_ {h'}$. Then, the equation for the difference is of the form investigated above, so that $h-h'=0$.
This means that there is a unique solution of the form
$M_\beta(v)
\rho(z)$ in the class of solutions with fixed density $\rho$. Then, putting this expression back in the equation we
determine $\rho$.
The next order equations for $\tilde f_i^{(n)}$ have a similar form, but the solutions are not anymore of
the form Maxwellian times a polynomial. The existence and uniqueness have to be  proved by a different
argument.

\bigskip

\section{\it Surface tension}
\setcounter{equation}{0}

The surface tension for a planar interface can be defined as the difference between the grand
canonical free energy (pressure)  of an equilibrium state with the interface and a homogeneous one
\cite{WR}. We can call excess
pressure this difference. The pressure for this model is
$${\mathcal P}(n_1,n_2) = \int {\rm d}x  p(n_1(x),n_2(x))$$
$$p(n_1,n_2)=T(n_1\log n_1 +n_2\log n_2) +
\frac{1}{2}n_1 U\star  n_2 +\frac{1}{2}n_2 U\star  n_1 -\mu_1
n_1-\mu_2n_2.$$
Consider the system in a cylinder of base $(2L)^{d-1}$ and height $M$  in presence of a planar interface
dividing the cylinder in the half upper cylinder where the densities are $ \rho_1^{+}, \rho_2^{+}$ and the
half lower cylinder with densities $
\rho_1^{-}, \rho_2^{-}$, where $ \rho_1^{\pm}, \rho_2^{\pm}$ are the equilibrium values of the densities in
the coexisting phases at temperature $T$. Then the excess pressure is given by 
\cite{B}  
$$\sigma=\lim_{L\to\infty}\frac{1}{(2L)^{d-1}}\lim_{M\to\infty}
\int_{-L}^L{\rm d}y_1\dots\int_{-L}^L{\rm d}y_{d-1}\int_{-M}^Mdy_d[p(w_1,w_2)-p( \rho_1^+, \rho_2^- )]$$
where
$w_i(q)$ are the front solutions, smooth functions satisfying the equations
\begin{equation}
T\log  w_i(q)+ \int _{\mathbb{R}} {\rm d}{q}'  \tilde U(|
{q}-{q}'|) w_j(q')=C_i
\label{three}
\end{equation}
where $\tilde U(q)= \int _{\mathbb{R} ^2} {\rm d}y U(\sqrt{q^{2}+y^{2}})$
and
$C_i=\mu_i-T$ are constants determined by the conditions at infinity $\rho_i^\pm$.  Notice that $f(
\rho_1^+,
\rho_2^-)=f( \rho_1^-, \rho_2^+ )$ since  $\rho^\pm_1=\rho_2^\mp$ and that $\mu_1=\mu_2=\mu$ in the
coexisting region.

We rewrite  the surface tension by using integration by part and the condition at
infinity
$$\sigma=\int_{-\infty}^{+\infty} {\rm d}z[p(w_1,w_2)-p( n_1^+, n_2^- )]=-\int_{-\infty}^{+\infty} {\rm
d}z z\frac{d}{dz}p(w_1(z),w_2(z)).$$
We have
$$\frac{d}{dz}p(w_1,w_2)=T[(\log w_1 +1)w'_1+ (\log w_2 +1)w'_2] +
\frac{1}{2}[w'_1 \tilde U\star  w_2 $$
$$+w'_2 \tilde U\star  w_1 +w_1 \tilde U\star  w'_2
+w_2 \tilde U\star  w'_1]-\mu(w_1'+w_2') .$$ By  using
 (\ref{three})  and $C_1=C_2=C$ we get
 $$\frac{d}{dz}f(w_1,w_2)=\frac{1}{2}[-w'_1 \tilde U\star  w_2 -w'_2 \tilde U\star  w_1 +w_1 \tilde U\star  w'_2
+w_2 \tilde U\star  w'_1] +(C+T)(w'_1 +w'_2)  -\mu(w_1'+w_2')$$ and for the surface tension, by remembering that $C=\mu-T$,
$$-\frac{1}{2}\int {\rm d}z {\rm d}z'\ z\sum_{i\neq j}[-w_i'(z)\tilde U(z-z')w_{j}(z')+w_{i}(z)\tilde
U(z-z')w_j'(z')]$$
In conclusion,
$$\sigma= \frac{1}{2}\int dz dz' (z-z')\sum_{i\neq j}[w^{'}_i(z)\tilde U(z-z')w_j(z')] .$$

\bigskip

\section{\it Forces}
\setcounter{equation}{0}

We show how to compute  the terms $\hat g_i^{(n)}$ and $\tilde g_i^{(n)}$ up to order 3. The procedure
can be easily extended at any order.

For a slowly varying function
$h( r,t)$ we have that
\begin{eqnarray}U^\e\star  h(r,t)&=&\int _{\mathbb{R}^{3}}\e^{-3}U(\e^{-1}|
{r}-{r}'|)h(r',t)dr'\cr&=&\int _{\mathbb{R}^{3}}U(|
x-x'|)[h(\e x',t)-h(\e x,t)]dx'+h(r,t)\int _{\mathbb{R}^{3}}U(|x-x'|)dx'\cr
&=& \int _{\mathbb{R}^{3}}U(|
x-x'|)\Big[\e(x-x')\cdot\nabla_r h(r,t)\cr
&+&\frac{\e^2}{2}\sum_{i,j}(x-x')_i(x-x')_j\frac{\partial^2}{\partial r_i\partial r_j} h(r,t)+\e^4R_h(x,x')\Big]dx'
+h(r,t)
\hat U\cr &=& h(r,t) \hat U+ \e^2 \Delta_r h(r,t) \bar U + \e^4U\star R_h
\label{Ubulk}
\end{eqnarray}
where $\hat U=\int U(r)dr,\quad \bar U=\frac{1}{2}\int r^2 U(r)dr$. We have used  the isotropy of $U$. Hence we
have $$\hat g_i^{(n)}= \hat U \hat \rho_i^{(n)},\  n=0,1; \quad\hat g_i^{(2)}= \hat U \hat \rho_i^{(2)} +\bar
U\Delta_r\hat \rho_i^{(0)},\quad \hat g_i^{(3)}=\hat U \hat\rho_i^{(3)} +\bar U\Delta_r\hat \rho_i^{(1)}. $$
\vskip.1cm To compute the expansion of $U^\e\star  h$ for a fast varying function $h(z,r,t)$ it is more convenient
to use a local system of coordinates. For a given curve $\Gamma$ and for any  point $s\in {\Gamma}$ we choose a
reference frame centered in $s$ with the axes $1,2$ along the directions of principal curvatures $k_i$ and $3$ in
the direction of the normal. Consider two points $r$ and $r'$ and choose the reference frame centered in $s(r):
r=s(r)+\e z\nu(r)$.  We denote  by $y_i$ and   $y'_i$ the microscopic coordinate of $r$ and $r'$ in this new
frame.  Then $y_1=y_2=0,y_3= z$ and $q'_i=\e^{-1}r_i'$, the microscopic coordinates of $r'$, are related to
$y'_i$ by a linear transformation $q'_i=A_{i\ell}y'_\ell$.
 Moreover, $z'$ is given in terms of $y'_i$
 by (\cite{GL})
\begin{eqnarray}
z'(\{y'_i\})&=y_3'+\sum_{i=1,2}\frac{1}{2}[\e k_i {y'_i}^2-2\e^2 k_i^2{y'}_i^2y'_3] +\frac{1}{2}\e^3
(\sum_ik_i^2{y'}_i^2)^2\cr
&-\e^3\frac{1}{4}[\sum_{ij}(
\partial_i^2k_j {y'_i}^2 {y'_j}^2(4k_i(1-\delta_{ij})+3-2\delta_{ij} )]  +O(\e^4),
\label{C.2}
\end{eqnarray}
where $z': r'=s(r')+ \e \nu(r') z'$.

We denote by $\check h(y'_1,y'_2,y'_3,t)$ the function $ h(\{\e A_{i\ell}y'_\ell\},z'(\{y'_i\}),t)$. We have
 \begin{eqnarray}
(U^\e\star  h)(z,r,t) &=&\int _{\mathbb{R}^{3}}dy'U(|y-y'|) \check h(y'_1,y'_2,,y'_3,t)
\cr
&=& \int _{\mathbb{R}^{3}}dy'U(|y-y'|) \check h( 0,0,y'_3, t)
+\frac{1}{2}\sum_{i=1,2}(\tilde U_{1,i}\star
\frac{\partial^2 \check h}{\partial {y'}_i^2})(z,r,t)\cr &+&\frac{1}{4}\sum_{i=1,2}(\tilde U_{2,i}\star
\frac{\partial^4 \check h}{\partial {y'}_i^4})(z,r,t)) +\frac{1}{4}\sum_{i,j=1,2, i\neq j}\tilde
U_{2,ij}\star
\frac{\partial^4 \check h}{\partial {y'}_i^2{y'}_j^2}+ Q,
\nonumber
\end{eqnarray}
 where

 $$\quad \tilde U_{s,i}(|y_3-y_3'|) =\int
_{\mathbb{R}^{2}}d y'_1dy'_2 U(\sqrt{|y_3-y_3'|^2+| y'_1|^2+| y'_2|^2}) |y'_i|^{2s}$$
$$\tilde U_{2,ij}=\int
_{\mathbb{R}^{2}}d  y'_1dy'_2 U(\sqrt{|y_3-y_3'|^2+| y'_1|^2+| y'_2|^2}) |y'_i|^2|y'_j|^2.$$
We have
$$\frac{\partial \check h}{\partial y'_i}=\e\sum_j\bar\nabla_jh A_{ji} +\frac{\partial h}{\partial z}
\frac{\partial z'}{\partial y'_i}$$
$$
\frac{\partial^2 \check h}{\partial {y'}_k^2}=\e^2\sum_{j\ell}A_{jk}A_{\ell k}\bar\nabla^2_{j\ell}h
+\frac{\partial z'}{\partial y'_k}[\e A_{jk}\bar\nabla_j\frac{\partial h}{\partial z}+\frac{\partial
z'}{\partial y'_k}\frac{\partial^2 h}{\partial z^2}]+\frac{\partial h}{\partial z}
\frac{\partial^2 z'}{\partial {y'}_k^2}.$$
By using the relation between $z'$ and $y_3'$  \ref{C.2}
 we see that the second term  equals to
 $$\frac{\partial h}{\partial
 z}(0,0,y'_3,t)(\e{k_k }-\e^2 2k_k^2 y'_3).$$
It is true that [B]
\begin{equation}
\int _{\mathbb{R}^{3}}dy' U(|y-y'|)\frac{\partial h}{\partial
 z}(0,0,y'_3,t)\sum_{i=1}^{d-1}\frac{k_i
^2{y'}_i^2}{2}= \frac{K}{2}\int _{\mathbb{R}}dz'(z'-z)\tilde U(|z'-z|)  h(z',r,t).
\label{presigma}
\end{equation}

To compute the contributions at different order in $\e$ we go back to the specific curve $\Gamma^\e_t$
and use the expansion $d^\e(r,t)=\sum_n \e^nd^{(n)}(r,t)$ which implies $k_i^\e=\sum_n \e^nk_i^{(n)}$
and
$A^\e_{ij}= \sum_n \e^nA_{ij}^{(n)}$. In conclusion,
\begin{eqnarray}
&&(U^\e\star  h)(z,r)=(\tilde U\star  h)(z,r) +\e\frac{\bar K}{2}\int _{\mathbb{R}}dz'(z'-z)\tilde
U(|z'-z|)  h(z',r) \cr &&+\sum_{i=1,2}\Big[\e^2\big(\tilde U_{1,i}
\star D_{1,i}(h)+\tilde U_{2,i}
\star D_{2,i}(h)+\sum_{j\neq i}\tilde U_{2,ij}
\star D_{2,ij}(h)+\frac{ K^{(1)}}{2} C(h)\big)+\e^3B_3\Big]\cr &&
:=(\tilde U\star  h)(z,r) +\sum_{n=1}^3\e^n B_n(h) +\tilde R_h,
\end{eqnarray}
where $\tilde R_h$ is of order $\e^4$ and
\goodbreak
$$ D_{1,i}(h)=\frac{1}{2}\sum_{j\ell}A^{(0)}{jk}A^{(0)}_{\ell i}\bar\nabla^2_{j\ell}h-\frac{\partial
h}{\partial
 z} 2(k^{(0)}_i)^2y'_3;\quad D_{2,i}(h)=
\frac{3-6y'_3}{4}(k^{(0)}_i)^2\frac{\partial^2 h}{\partial z^2};$$
$$C(h)=\int _{\mathbb{R}}dz'(z'-z)\tilde
U(|z'-z|)  h(z',r),\quad D_{2,ij}(h)=\frac{1}{4}k^{(0)}_ik^{(0)}_j\frac{\partial^2 h}{\partial z^2}.$$
We do not write explicitly the long and uninteresting formula for  $B_3$. Hence we have
\begin{eqnarray}
&&\tilde g_i^{(0)}= \tilde U \star \tilde \rho_i^{(0)},\cr &&
\tilde g_i^{(1)}= \tilde U\star
\tilde\rho_i^{(1)} +\e B_1 (\tilde \rho_i^{(0)})(z,r)\cr &&
\tilde g_i^{(2)}= \tilde U \star \tilde \rho_i^{(2)} + B_1(\tilde \rho_i^{(1)}) +
B_2(\tilde \rho_i^{(0)})\cr &&
\tilde g_i^{(3)}=\tilde U \star \tilde \rho_i^{(3)} +B_1(\tilde \rho_i^{(2)}) +
B_2(\tilde \rho_i^{(1)})+ B_3(\tilde \rho_i^{(0)}).
\end{eqnarray}

\bigskip

\section*{Acknowledgements}
R. M. is indebted to M. C. Carvalho and E. Carlen for enlightening discussions in the initial phase of
this project and to R. Esposito for many valuable discussions. R. M. likes to thank E. Orlandi  for
kindly sending her the draft of the paper \cite{CCO1} as well as for interesting discussions.

\end{document}